\RequirePackage{fix-cm}
\documentclass[twocolumn,epjc3]{svjour3} 
\input{preamble.tex}

\journalname{Eur. Phys. J. C}

\begin{document}
\sloppy 
\title{Identification of the cosmogenic \Cel background in large volumes of liquid scintillators with Borexino} 
\def\APC{AstroParticule et Cosmologie, Universit\'e Paris Diderot, CNRS/IN2P3, CEA/IRFU, Observatoire de Paris, Sorbonne Paris Cit\'e, 75205 Paris Cedex 13, France}

\def\Dubna{Joint Institute for Nuclear Research, 141980 Dubna, Russia}
\def\Genova{Dipartimento di Fisica, Universit\`a degli Studi e INFN, 16146 Genova, Italy}
\def\Krakow{M.~Smoluchowski Institute of Physics, Jagiellonian University, 30059 Krakow, Poland}
\def\Kiev{Kiev Institute for Nuclear Research, 03680 Kiev, Ukraine}
\def\Kurchatov{National Research Centre Kurchatov Institute, 123182 Moscow, Russia}
\def\Kurchatovb{ National Research Nuclear University MEPhI (Moscow Engineering Physics Institute), 115409 Moscow, Russia}
\def\LNGS{INFN Laboratori Nazionali del Gran Sasso, 67010 Assergi (AQ), Italy}
\def\Milano{Dipartimento di Fisica, Universit\`a degli Studi e INFN, 20133 Milano, Italy}
\def\Perugia{Dipartimento di Chimica, Biologia e Biotecnologie, Universit\`a degli Studi e INFN, 06123 Perugia, Italy}
\def\Peters{St. Petersburg Nuclear Physics Institute NRC Kurchatov Institute, 188350 Gatchina, Russia}
\def\Princeton{Physics Department, Princeton University, Princeton, NJ 08544, USA}
\def\PrincetonChemEng{Chemical Engineering Department, Princeton University, Princeton, NJ 08544, USA}
\def\UMass{Amherst Center for Fundamental Interactions and Physics Department, University of Massachusetts, Amherst, MA 01003, USA}
\def\Virginia{Physics Department, Virginia Polytechnic Institute and State University, Blacksburg, VA 24061, USA}
\def\Munchen{Physik-Department and Excellence Cluster Universe, Technische Universit\"at  M\"unchen, 85748 Garching, Germany}
\def\Lomonosov{ Lomonosov Moscow State University Skobeltsyn Institute of Nuclear Physics, 119234 Moscow, Russia}
\def\GSSI{ Gran Sasso Science Institute (INFN), 67100 L'Aquila, Italy}
\def\Dresda{Department of Physics, Technische Universit\"at Dresden, 01062 Dresden, Germany}
\def\Mainz{Institute of Physics and Excellence Cluster PRISMA$^+$, Johannes Gutenberg-Universit\"at Mainz, 55099 Mainz, Germany}
\def\Juelich{Institute for Nuclear Physics IKP-2, Forschungszentrum J\"ulich, 52428 J\"ulich, Germany}
\def\RWTH{Physics Institute IIIB, RWTH Aachen University, 52062 Aachen, Germany}
\def\London{Department of Physics, Royal Holloway, University of London, Department of Physics, School of Engineering, Physical and
Mathematical Sciences, Egham, Surrey, TW20 OEX, UK}
\def\LondonUCL{Department of Physics and Astronomy, University College London, London, UK}
\def\Atomki{Institute of Nuclear Research (Atomki), Debrecen, Hungary}

\def\Berkeley{University of California, Berkeley, Department of Physics, CA 94720, Berkeley, USA}
\def\Padova{Dipartimento di Fisica e Astronomia dell'Universit\`a di Padova and INFN Sezione di Padova, Padova, Italy}
\def\Napoli{Dipartimento di Fisica, Universit\`a degli Studi Federico II e INFN, 80126 Napoli, Italy}
\def\UGent{Departement Fysica en Sterrenkunde, Universiteit Gent, Gent, Belgium}
\def\Madrid{Universidad Aut\'onoma de Madrid, Ciudad Universitaria de Cantoblanco, 28049 Madrid, Spain}

\author{
    M.~Agostini\thanksref{33, 3} \and
	K.~Altenm\"{u}ller\thanksref{3} \and
	S.~Appel\thanksref{3} \and
	V.~Atroshchenko\thanksref{4} \and
	Z.~Bagdasarian\thanksref{5, 36} \and
	D.~Basilico\thanksref{7} \and
	G.~Bellini\thanksref{7} \and
	J.~Benziger\thanksref{8} \and
	R.~Biondi\thanksref{10} \and
	D.~Bravo\thanksref{7,37} \and
	B.~Caccianiga\thanksref{7} \and
	F.~Calaprice\thanksref{12} \and
	A.~Caminata\thanksref{13} \and
	P.~Cavalcante\thanksref{11, 10} \and
	A.~Chepurnov\thanksref{14} \and
	D.~D'Angelo\thanksref{7} \and
	S.~Davini\thanksref{13} \and
	A.~Derbin\thanksref{17} \and
    A. DiZGiacintio\thanksref{10} \and
    V. Di~Marcello\thanksref{10} \and
	X.F.~Ding\thanksref{12} \and
	A.~Di~Ludovico\thanksref{12} \and
	L.~Di~Noto\thanksref{13} \and
	I.~Drachnev\thanksref{17} \and
	A.~Formozov\thanksref{7, 18} \and
	D.~Franco\thanksref{19} \and
	C.~Galbiati\thanksref{12,2} \and
	C.~Ghiano\thanksref{10} \and
	M.~Giammarchi\thanksref{7} \and
	A.~Goretti\thanksref{12, 10} \and
    A.S.~G\"ottel\thanksref{5,6} \and
    M.~Gromov\thanksref{14, 18} \and
	D.~Guffanti\thanksref{1} \and
	Aldo~Ianni\thanksref{10} \and
	Andrea~Ianni\thanksref{12} \and
	A.~Jany\thanksref{24} \and
	D.~Jeschke\thanksref{3} \and
	V.~Kobychev\thanksref{25} \and
	G.~Korga\thanksref{32,39} \and
	S.~Kumaran\thanksref{5,6} \and
	M.~Laubenstein\thanksref{10} \and
	E.~Litvinovich\thanksref{4,26} \and
	P.~Lombardi\thanksref{7} \and
	I.~Lomskaya\thanksref{17} \and
	L.~Ludhova\thanksref{5,6} \and
	G.~Lukyanchenko\thanksref{4} \and
	L.~Lukyanchenko\thanksref{4} \and
	I.~Machulin\thanksref{4,26} \and
	J.~Martyn\thanksref{1} \and
	E.~Meroni\thanksref{7} \and
	M.~Meyer\thanksref{28} \and
	L.~Miramonti\thanksref{7} \and
	M.~Misiaszek\thanksref{24} \and
	V.~Muratova\thanksref{17} \and
	B.~Neumair\thanksref{3} \and
	M.~Nieslony\thanksref{1} \and
	R.~Nugmanov\thanksref{4,26} \and
	L.~Oberauer\thanksref{3} \and
	V.~Orekhov\thanksref{1} \and
	F.~Ortica\thanksref{29} \and
	M.~Pallavicini\thanksref{13} \and
	L.~Papp\thanksref{3} \and
	L. Pelicci\thanksref{5,6} \and
	\"O.~Penek\thanksref{5} \and
	L.~Pietrofaccia\thanksref{12} \and
	N.~Pilipenko\thanksref{17} \and
	A.~Pocar\thanksref{30} \and
	A.~Porcelli\thanksref{1,38}\and
	G.~Raikov\thanksref{4} \and
	M.T.~Ranalli\thanksref{10} \and
	G.~Ranucci\thanksref{7} \and
	A.~Razeto\thanksref{10} \and
	A.~Re\thanksref{7} \and
	M.~\mbox{Redchuk \thanksref{5,6,34}} \and
	A.~Romani\thanksref{29} \and
	N.~Rossi\thanksref{10} \and
	S.~Sch\"onert\thanksref{3} \and
	D.~Semenov\thanksref{17} \and
	G.~Settanta\thanksref{5} \and
	M.~Skorokhvatov\thanksref{4,26} \and
	A.~Singhal\thanksref{5,6} \and
	O.~Smirnov\thanksref{18} \and
	A.~Sotnikov\thanksref{18} \and
	Y.~Suvorov\thanksref{4,10,35} \and
	R.~Tartaglia\thanksref{10} \and
	G.~Testera\thanksref{13} \and
	J.~Thurn\thanksref{28} \and
	E.~Unzhakov\thanksref{17} \and
	A.~Vishneva\thanksref{18} \and
	R.B.~Vogelaar\thanksref{11} \and
	F.~von~Feilitzsch\thanksref{3} \and
	M.~Wojcik\thanksref{24} \and
	M.~Wurm\thanksref{1} \and
	S.~Zavatarelli\thanksref{13} \and
	K.~Zuber\thanksref{28} \and
	G.~Zuzel\thanksref{24}
}

\institute{
		\LondonUCL\label{33}\and
		\Munchen \label{3}\and
		\Kurchatov \label{4}\and
		\Juelich \label{5}\and
		\Milano \label{7}\and
		\PrincetonChemEng \label{8}\and
		\LNGS \label{10}\and
		\Princeton \label{12}\and
		\Genova \label{13}\and
		\Virginia \label{11}\and
		\Lomonosov \label{14}\and
		\Peters \label{17}\and
		\Dubna \label{18}\and
		\APC \label{19}\and
		\GSSI \label{2}\and
		\RWTH \label{6}\and
		\Mainz\label{1} \and
		\Krakow \label{24}\and
		\Kiev \label{25}\and
		\London \label{32}\and
		\Atomki\label{39}\and
		\Kurchatovb \label{26}\and
		\Dresda \label{28}\and
		\Perugia \label{29}\and
		\UMass \label{30}\and
		\emph{Present Address:} \Berkeley\label{36}\and
		\emph{Present Address:} \Madrid\label{37}\and
		\emph{Present Address:} \UGent\label{38}\and
		\emph{Present Address:} \Padova\label{34}\and
		\emph{Present Address:} \Napoli\label{35}
}

\onecolumn
\maketitle
\twocolumn


\begin{abstract}
	Cosmogenic radio-nuclei are an important source of background for low-energy neutrino experiments. 
In Borexino, cosmogenic \Cel decays outnumber solar $pep$ and CNO neutrino events by about ten to one. 
In order to extract the flux of these two neutrino species, a highly efficient identification of this background is mandatory.
We present here the details of the most consolidated strategy, used throughout Borexino solar neutrino measurements. 
It hinges upon finding the space-time correlations between \Cel decays, the preceding parent muons and the accompanying neutrons.
This article describes the working principles and evaluates the performance of this Three-Fold Coincidence (TFC) technique in its two current implementations: a hard-cut and a likelihood-based approach.
Both show stable performances throughout Borexino Phases II (2012-2016)  and III (2016-2020) data sets, with a \Cel tagging efficiency of $\sim$90\,\% and $\sim$63-66\,\% of the exposure surviving the tagging. 
We present also a novel technique that targets specifically \Cel produced in high-multiplicity during major spallation events.
Such \Cel appear as a {\it burst} of events, whose space-time correlation can be exploited.   
Burst identification can be combined with the TFC to obtain about the same tagging efficiency of $\sim$90\% but with a higher fraction of the exposure surviving, in the range of $\sim$66-68\,\%.
\end{abstract}


\section{Introduction}\label{sec:intro}
	Since 2007, the Borexino detector has been observing solar neutrinos from the depths of the Gran Sasso National Laboratory (Italy). 
Designed for the spectroscopy of low energy neutrinos detected via elastic scattering off electrons, Borexino has accomplished individual measurements of the solar $^7$Be \cite{bx_7Be}, $^8$B \cite{bx_8B}, $pep$ \cite{bx_pep}, and $pp$ neutrino fluxes \cite{bx_pp}. 
Moreover, Borexino has performed a comprehensive spectroscopic measurement of the entire neutrino spectrum emitted in the $pp$-chain \cite{bx_nature, bx_nusol, bx_8B_new}. 
Most recently, Borexino achieved the first direct observation of neutrinos produced in the solar CNO cycle  \cite{bx_cno, bx_cno_sens}.  

While there are several elements that contributed to the success of Borexino, we focus here on the analysis strategy that has been developed for the identification of cosmogenic background in the 1--2~MeV spectral region (originally proposed in \cite{tfc_deutsch}).
The observation of $pep$ and CNO neutrinos is made possible by the suppression of the $\beta^+$-decays of \Cel that are otherwise largely indistinguishable from neutrino-induced electron recoils on an event-by-event basis.
In organic scintillators, \Cel is being produced by cosmic muons that penetrate the rock coverage of the laboratory and induce spallation processes on carbon of type $\mu+^{12}\text{C}\rightarrow \mu+^{11}\text{C}+n\label{eq:C11_prod}$. 
A discussion of the processes that contribute to \Cel isotope production can be found in \cite{c11prod}.
The isotope features a lifetime of 29.4~minutes and a $Q$-value of 0.96~MeV. Due to positron annihilation, the visible spectrum is shifted to higher energies, covering a range between $\sim$0.8 and $\sim$2~MeV. 

\begin{figure}
\centering
\includegraphics[width=\linewidth]{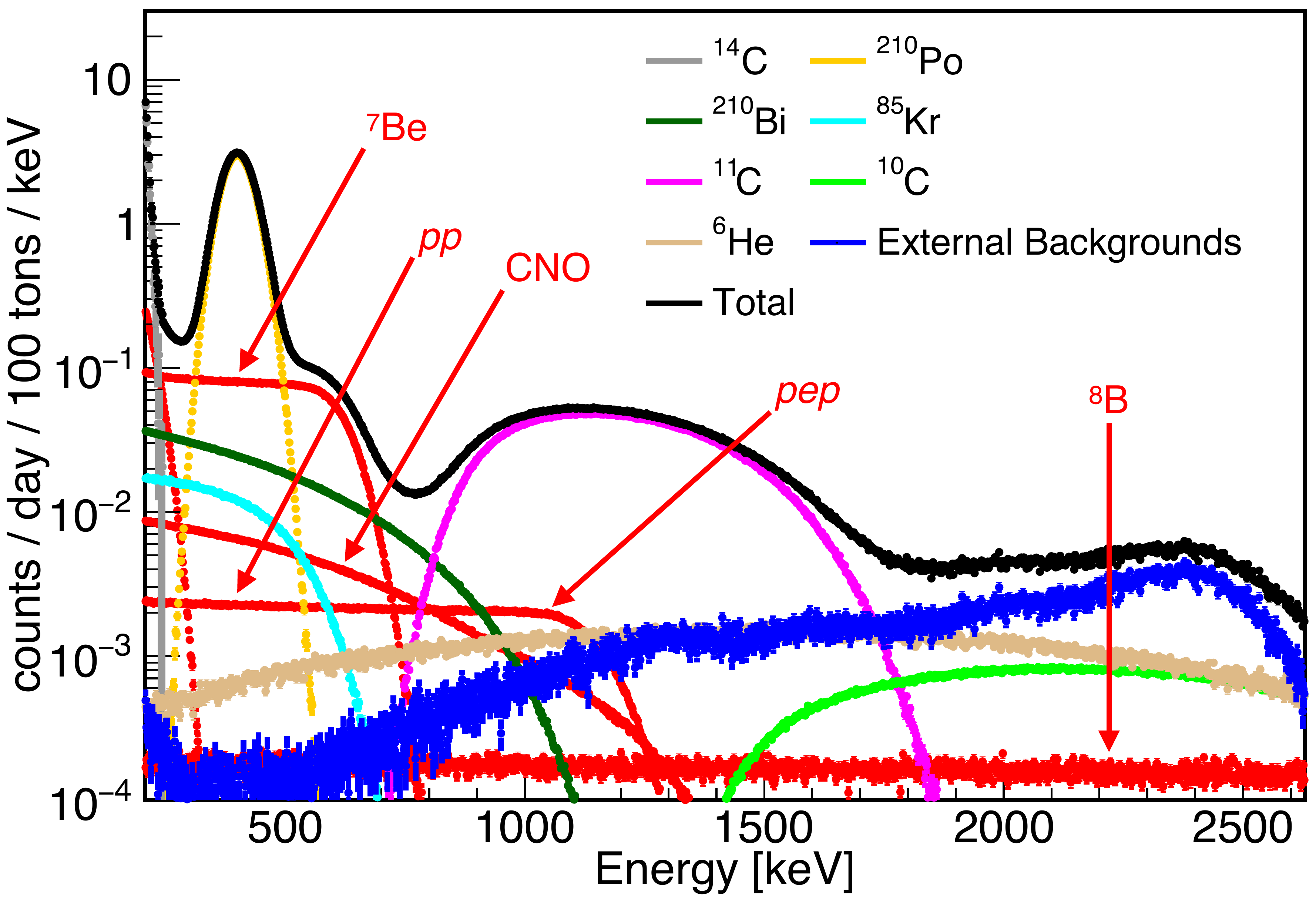}
\caption{Expected spectrum in Borexino from Monte Carlo simulations. 
Electron recoil energy due to neutrino interactions (red lines) and  background components (other colors). The Fiducial Volume applied is described in \cref{sec:borexino}.}
\label{fig:mc_spectrum}
\end{figure}

\cref{fig:mc_spectrum} shows the main neutrino signal and radioactive background contributions to the visible energy spectrum of Borexino. At first glance, detection of $pep$ and CNO neutrinos seems not very promising since other solar neutrino species cover the lower part of their spectra while towards the upper end \Cel overtops their signals by about an order of magnitude. An efficient veto strategy for \Cel is a necessary prerequisite for overcoming this disadvantageous situation.

The present article illustrates two analysis techniques for an effective \Cel suppression. 
The first is the Three-Fold Coincidence (TFC) technique. 
The TFC is formed by the space and time correlation of \Cel events with their parent muons and the neutrons that in most cases accompany \Cel production. 
By now, the technique has been extensively used in Borexino analyses. 
TFC creates two subsets of data: (1) a \Cel-depleted set that is including over 60\% of the exposure but less than 10\% of the \Cel events; (2) a \Cel-enriched set with the complementary exposure and \Cel events. 
Both datasets are used in the multivariate fit that leads to the extraction of the solar neutrino signals as explained in~\cite{bx_long, bx_nusol}.

The second technique is the Burst Identification (BI) technique, a new tagging strategy that targets \Cel events originating from a single hadronic shower upon the passage of a muon. 
These events are visible in {\it bursts} and are space-time correlated among themselves: due to the negligible convective motion of the scintillator in the timescale of the \Cel mean life, they remain essentially aligned along the parent muon track. 
Since the BI technique on its own does not provide sufficient tagging efficiency for solar neutrino analyses, it is only used in combination with the TFC in Borexino. 
However, the BI method might be interesting as a stand-alone solution for experimental setups that do not provide sufficient information for a TFC-like approach.

In this article, we first recall the key features of the Borexino detector (\cref{sec:borexino}); we then review the TFC technique, presenting the two most developed implementations that use either a cut-based or likelihood-based approach and their comparative performances (\cref{sec:tfc}); we go on to discuss the new BI technique and the gain in \Cel-depleted exposure that can result from a combined application of TFC and BI (\cref{sec:bursts}), before presenting our conclusions in \cref{sec:conclusions}.

It is worth to note that both the TFC and the BI algorithms have been developed for the tagging of \Cel but are not limited to this specific cosmogenic isotope. 
Despite some variability in the underlying production processes \cite{kamland_cosmogenics}, the creation of radioactive isotopes in organic scintillator is closely linked to muons inducing hadronic showers, resulting in especially bright muon events and subsequent {\it bursts} of delayed neutron captures. 
Both tagging techniques are of immediate use for upcoming neutrino detectors based on organic scintillators \cite{snoplus, juno} but may as well be of interest for water Cherenkov detectors with neutron-tagging capabilities \cite{gadzooks, HK,SK_tagging} or hybrid detectors \cite{theia}. 

\section{The Borexino detector}\label{sec:borexino}
	\begin{figure*}
\centering
\includegraphics[width=0.7\linewidth]{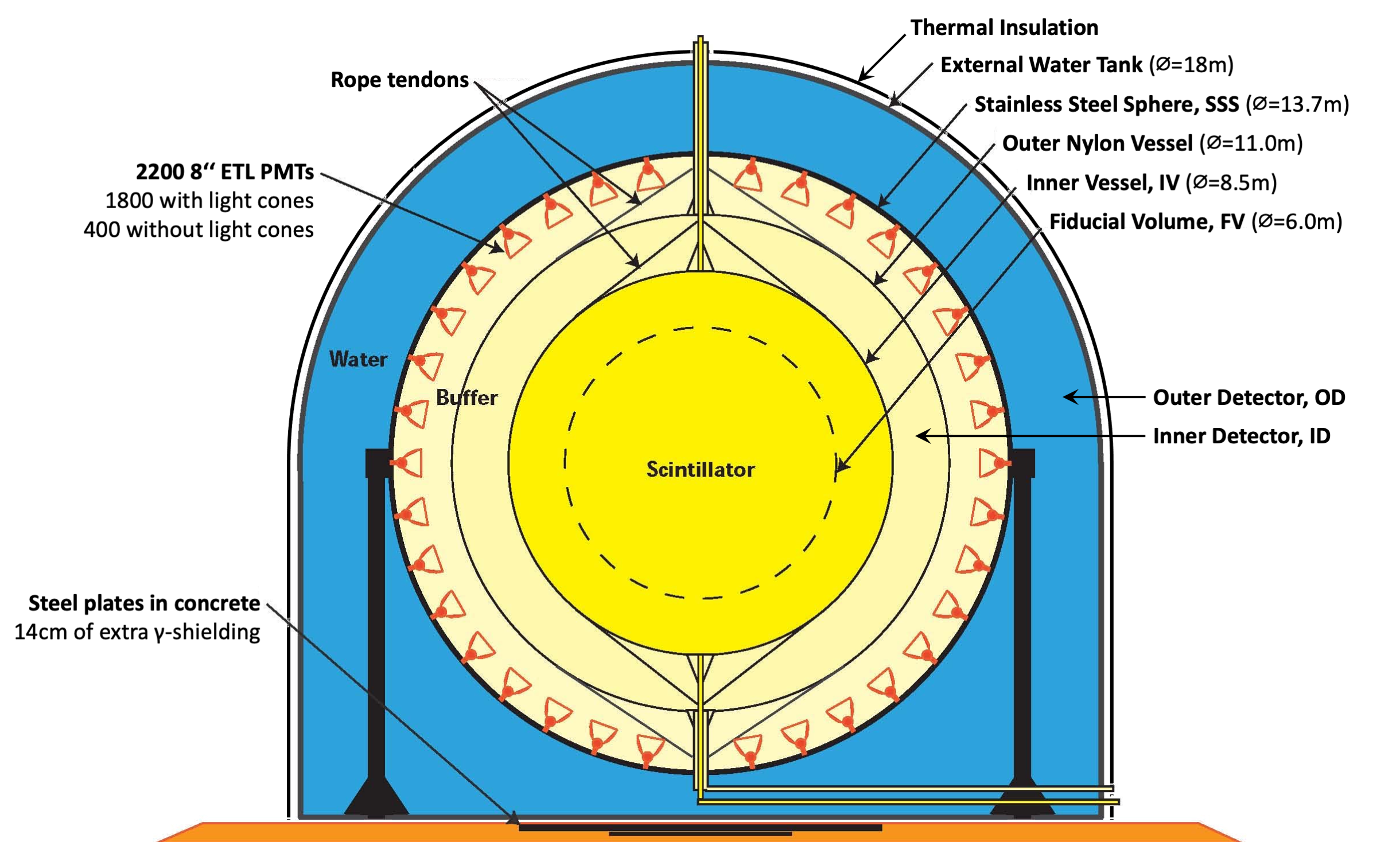}
\caption{Sketch of the Borexino Detector.}
\label{fig:borex}
\end{figure*}

A schematic drawing of the Borexino detector \cite{bx_detector} is shown in \cref{fig:borex}. 
The neutrino target consists of 278\,t of organic scintillator composed of the solvent PC (1,2,4-trimethylbenzene) doped with the wavelength shifter PPO (2,5-diphenyloxazole) at a concentration of 1.5\,g/l. 
The scintillator mixture is contained in a spherical and transparent nylon Inner Vessel (IV) with a diameter of 8.5\,m and a thickness of 125\,$\upmu$m. 
To shield this central target from external $\gamma$-ray backgrounds and to absorb emanating radon, the IV is surrounded by two layers of buffer liquid in which the light quencher DMP (dimethylphthalate) is added to the scintillator solvent. 
A Stainless Steel Sphere (SSS) of 13.7\,m diameter holding 2212 8'' photomultiplier tubes (PMTs) completes the Inner Detector (ID). 
PMTs are inward-facing and detect the scintillation light caused by particle interactions in the central region.
The ID is embedded in a steel dome of 18\,m diameter and 16.9\,m height that is filled with 2.1\,kt of ultra-pure water.
The outer surface of the SSS and the floor of the water tank are instrumented with additional 208 PMTs. 
This Outer Detector (OD) provides efficient detection and tracking of cosmic muons via the Cherenkov light that is emitted during their passage through the water.
Neutrons created by the passage of muons are quickly thermalized and captured on hydrogen (carbon in 1\% of the cases) in the scintillator after $\tau_n\sim260\,\upmu$s \cite{bx_cosmogenics} with the emission of a  2.2\,MeV (4.9\,MeV) de-excitation gamma-ray. 
Muon and neutron detection in Borexino is described in detail in~\cite{bx_od}. 
A thorough analysis of all cosmogenic backgrounds can be found in~\cite{bx_cosmogenics} and the most recent update on the cosmic muons and neutrons in~\cite{bx_muon_seasonal}.

\paragraph{Event reconstruction.} In Borexino, a physical event is defined by the firing of several PMTs of the ID within a short period of time (typically 20--25 PMTs in 60--100\,ns, changing over time).  This occurrence causes a trigger signal and opens an acquisition window of 16~$\upmu$s during which all PMT pulses are saved. 
As this window may contain more than one physical event, an offline \emph{clustering} algorithm identifies pulses within a short time window (several 100\,ns) as belonging to a specific event. 
Since the scintillation light output is proportional to the deposited energy, the number of pulses (``hit'') recorded by the PMTs for a given \emph{cluster}, $N_\text{hits}$, is used as energy estimator. We record about 500 photoelectrons per MeV of deposited energy. 
In order to account for the increasing number of malfunctioning PMTs throughout the data set~\cite{bx_nusol}, the number of hits is always normalized to 2000 live channels. The position of the scintillation events in the scintillator volume is reconstructed based on the time-of-flight of the scintillation photons from vertex to PMTs. The reconstruction algorithm uses PDFs for the expected arrival time distributions at the PMTs to take into account both propagation effects and the time profile of the scintillation process \cite{bx_long}.

\paragraph{Muon and Neutron Windows.} 
A muon crossing the scintillator target is expected to cause sizable signals in both ID and OD\footnote{Muons crossing only the OD instead do not play a role in this analysis.}.
As for ordinary events, the ID trigger opens a 16~$\upmu$s-long ``Main Window'' (MW) for which all hits are saved. 
The OD issues an additional trigger flag with high efficiency.
In case of its presence, the MW is followed by a substantially longer ``Neutron Window'' (NW): 1.6\,ms. 
This is meant to catch the subsequent signals due to the capture of muon-induced neutrons without requiring explicit triggers. 
Given the value of $\tau_n$, about 6\% of neutrons are collected in the MW and the rest in the NW. 

\paragraph{Empty Boards condition.} The ID is designed for low energy spectroscopy, so scintillator-crossing muons that generate a huge amount of light have a significant chance of saturating the electronics memory. 
Consequently, the reconstruction of the energy and of the position of the neutron-capture gammas following such event can be troublesome. Effects are more severe in the MW than in the NW.
It has been found that this condition can be quantified by the number of electronic boards that show no data for MW and NW in any of their eight channels when memory saturation occurs. Thus, the number of boards devoid of hits (out of a total of 280 boards) is counted and stored for analysis as parameter $N_\text{EB}$ (\# of Empty Boards). 

\paragraph{Neutron identification.} Luminous muon events produce a long tail of trailing hits that extend into the NW (e.g.~by after pulsing of PMTs). A specifically adapted \emph{clustering} algorithm has been developed for both the muon MW and NW to find physical events superimposed to this background. This task is further complicated be the aforementioned presence of saturation effects.

Due to its considerable length the NW contains not only neutrons (and cosmogenic isotopes) but for the most part accidental counts. 
These are dominated by low energy \A{C}{14} decays that occur at a rate of about 100 s$^{-1}$ \cite{bx_pp}. 
To avoid mis-counting those and other low-energy events as neutron captures, an energy selection threshold has been introduced at $N_\text{thr}=385$~hits (corresponding roughly to 770\,keV). 
To take into account the saturation effects that diminish the number of hits collected from a neutron event, 
the selection criterium is adjusted to $N_\text{hits} > N_\text{thr} - 2\times N_\text{EB}$.

\paragraph{Neutron vertex reconstruction.} Following bright muons, the accuracy of vertex reconstruction can be severely affected by electronics saturation effects.
Therefore, position reconstruction is ineffective for neutrons in the MW and can still be severely affected in the NW.
The TFC algorithms described below take this into account by introducing Full Volume Vetoes (see \cref{sec:blind_cuts}) in presence of severe saturation effects.
The BI algorithm described in \cref{sec:bursts} can be used as an alternative, more effective way to target \Cel produced under these circumstances.

\paragraph{Fiducial volume.} Unless otherwise noted, in this paper we refer to fiducial volume as to the one used in the $pp$-chain \cite{bx_nature} and in the CNO \cite{bx_cno} neutrino analyses.
This means considering events with a reconstructed position within a 2.8~m radius from the center of the detector and vertical coordinate in the range [-1.8, 2.2]~m.

\paragraph{Data taking campaigns.} The Borexino data used throughout this paper refer to Phase-II and Phase-III of the data taking campaign, ranging respectively from 14 December 2011 till 21 May 2016 and from 17 July 2016 till 1 March 2020. The average number of working PMTs during Phase-II and III was about 1500 and about 1200, respectively.

A temperature control strategy was gradually adopted throughout Phase-III to prevent convection in the liquid scintillator \cite{bx_cno} in order to stabilize internal radioactive background levels. However we note that no impact is expected in \Cel suppression mechanisms as the observed convective time scale is much longer than \Cel mean life.

\section{Three-Fold Coincidence}\label{sec:tfc}
	The TFC identifies \Cel candidates by their space and time correlation with the preceding parent muons and muon-induced neutrons
and splits the original data sample into two sub-samples: one containing the \Cel-tagged events (enriched sample) and another one where most \Cel events are removed (depleted sample).

In practice, the procedure is complicated by the relatively long lifetime of the \Cel isotope ($\tau_{^{11}{\rm C}} =29.4$\,min) compared to the rate of muons crossing the Borexino ID (3\,min$^{-1}$). 
While the association of muon tracks and subsequent neutron capture vertices ($\tau_n\sim260\,\upmu$s) is straight-forward, tagging conditions have to be maintained for several $\tau_{^{11}{\rm C}}$ and thus select not only \Cel candidates but also substantial amounts of non-cosmogenic events. 
While we refer to the latter as \Neu events in the following, it is worth to point out that this event category contains as well a large amount of decays caused by internal and external radioactivity that we include under this denomination for the sake of simplicity. 

Typically, about 40~\% of the exposure is contained in the \Cel-enriched data set that includes over 90\% of \Cel decays. The complementary \Cel-depleted data set thus mostly consists of {\it neutrino} events and corresponds to $\sim$60\,\% of the exposure. 
This latter set is the most relevant for $pep$ and CNO neutrino analyses\footnote{Nevertheless, Borexino multivariate fit strategy uses both subsets as explained in \cite{bx_long, bx_nusol}. This maximises the sensitivity to determine neutrino and background signals outside the \Cel energy region.}. 
Thus, the goal of the TFC algorithms is to minimize the \Cel that remains in the \Cel-depleted subset and to maximize the fraction of the exposure that contributes to it.

We note that in \cite{c11prod} it is theoretically calculated that about 5\% of the \Cel production does not have a neutron associated. Although in principle this could imply an inefficiency of the TFC method, several \Cel atoms and additional neutrons are generally produced by the same parent muon, effectively limiting the impact of these invisible channels. 

This section describes the two TFC approaches used in Borexino analyses: the Hard Cut approach (HC-TFC, \cref{sec:mi_tfc}) and the Likelihood approach (LH-TFC, \cref{sec:mz_tfc}). 

HC-TFC generates a list of space-time veto regions within the exposure upon meeting certain conditions (typically a muon followed by one or more neutrons).
All regular events are then tested against the regions in the list and if they fall within one (or more) they are labelled as \Cel. 
LH-TFC builds for each event a likelihood to be a \Cel decay based of its space-time distance from preceding muon-neutron pairs and applies the \Cel tag if the likelihood exceeds a programmable threshold.
\Cel-tagged events will include a fraction of falsely tagged \Neu events that are thus included in the  \Cel-enriched spectrum. Vice versa, the \Cel-depleted data set will contain a residual of \Cel decays. 
Their fraction represents the inefficiency of the TFC method.

\begin{figure}[t]
	\centering
	\includegraphics[width=\linewidth]{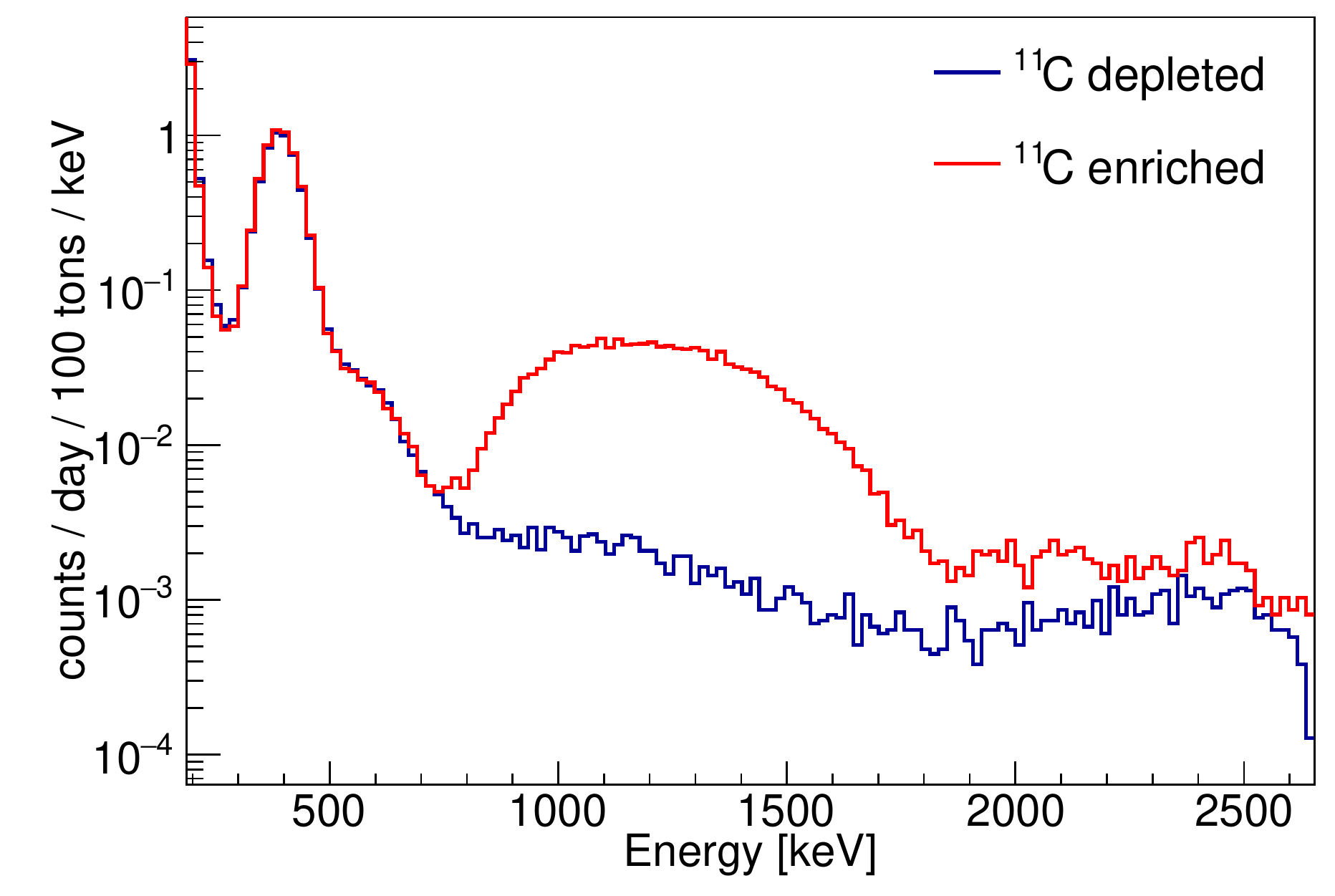}
	\caption{\Cel-depleted (blue line) and enriched (red line) charge spectra created with the TFC method for Phase-II. Spectra are normalized to the same exposure. Only basic cuts that remove muon and noise events are applied. The spectral difference above 1800\,keV is caused by the TFC tagging of cosmogenic $^{6}$He and $^{10}$C.}
	\label{fig:btfc_spectrums_compare}
\end{figure}

\cref{fig:btfc_spectrums_compare} shows the \Cel-depleted and \mbox{-enriched} visible energy spectra based on Borexino Phase-II data. 
The effect of the TFC in the \Cel energy range (0.75--1.87\,MeV) is striking. 
A difference in the spectral shapes persists at even higher energies. 
This difference can be attributed to other cosmogenic isotopes (predominantly \A{C}{10} and \A{He}{6}) that feature higher spectral end-points and that are picked up as well by the TFC tagging.

The fraction of exposure that contributes to each spectrum is computed within each TFC algorithm by generating random events with uniform time and position distributions within the fiducial volume, effectively reproducing every real run. These events are then tested against the same list of veto regions or likelihood condition of the real events.

	\subsection{Hard Cut approach}\label{sec:mi_tfc}
		The first approach for the suppression of the \Cel background developed in Borexino \cite{bx_long} is based on defining a list of space-time veto regions in the dataset.
A region is added  to the list upon the occurrence of a combination of muon and neutron events.
Each region holds for a programmable time after the events that generated it. We set this time to 2.5\,h, approximately five times the \Cel mean time, based on an optimization for tagging efficiency and exposure (see \cref{sec:comparisons}).
Regular events are then checked against the list of regions and if falling within one or more they are flagged as \Cel.

We now describe the different conditions that determine the creation of a veto region. 
In the definition of the regions we use the concept of neutron multiplicity, indicating how many clusters of PMT hits that follow a muon event could possibly be due to a $n$-capture gamma. 
In the MW we count $n_{\mu n}$ as the number of clusters found by the ad-hoc cluster-finding algorithm that accounts for the underlying tail of the muon event, excluding the first 3.5\,$\upmu$s of the window. 
In the NW we count $n_{n}$ as the number of clusters that meet the requirement expressed in \cref{sec:borexino}. 

\begin{figure}
\centering
\includegraphics[width=1\linewidth]{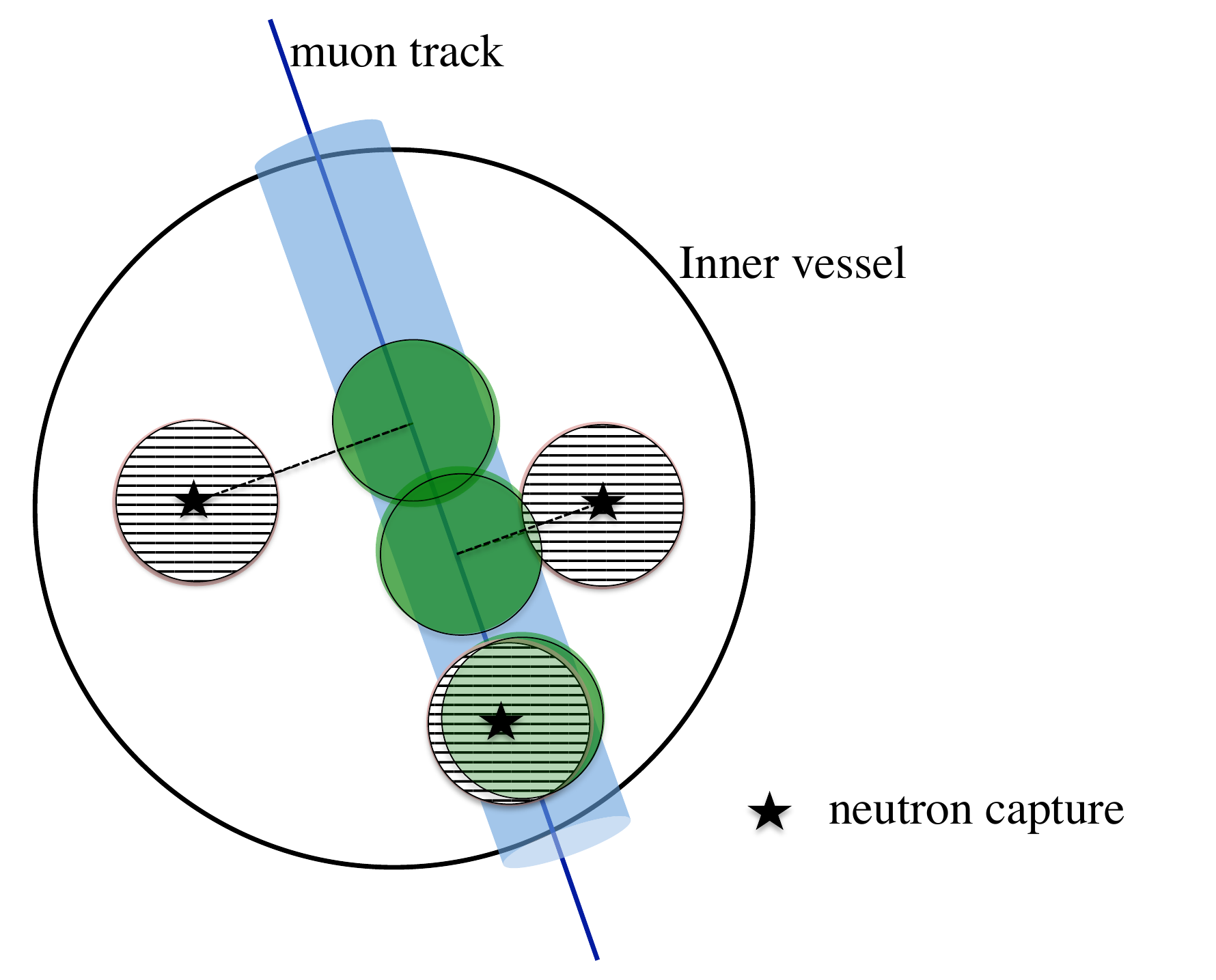}
\caption{HC-TFC working principle. Upon the passage of a muon through the IV, geometrical veto regions are defined: a cylinder around the muon track (blue), and spheres around the n-capture points (shaded) and around their projection upon the track (green). The typical radius of the cylindrical (spherical) regions is 0.7~m (1.2~m). The picture is not to scale.}
\label{fig:tfc_veto}
\end{figure}

Upon finding a muon followed by at least one neutron cluster, we define geometrical veto regions: (1) a cylindrical region around the reconstructed muon track (if available); (2) a spherical region around the $n$-capture reconstructed position in the NW; (3) a spherical region around the projection point of the $n$-capture position in the NW onto the muon track (if available). 
The working principle is depicted in \cref{fig:tfc_veto}.
After tuning for tagging efficiency and exposure (see \cref{sec:comparisons}), the radius of the cylindrical and spherical regions are chosen to be 0.7\,m and 1.2\,m, respectively as summarized in \cref{fig:tfc_veto} caption.

Concerning the tracking of muons, as reported in \cite{bx_od}, Borexino has developed two algorithms based on the ID signals and one based on the Cherenkov light recorded by the OD.
The track resulting from a global fit of four entry/exit-points (two from either of the ID-based algorithms and two from the OD-based one) is chosen for the vast majority of muons. 
In those cases when the fit result is inadequate, we rely on the available results of the individual algorithms \cite{bx_od}. In the rare cases ($\sim$\,0.1\,$\mu$/d) for which none of the tracking algorithms could provide a meaningful track, we apply a Full Volume Veto (introduced in \cref{sec:blind_cuts}).

In the years up to 2012, the CNGS neutrino beam aimed from CERN to Gran Sasso induced $\nu_\mu$ interactions in both the Borexino detector and the rock upstream, creating a sizable amount of muons crossing the scintillator volume  \cite{bx_cngs}. 
These muons are relatively low in energy ($\sim$\,20\,GeV) and, as expected, we observe that they do not significantly contribute to \Cel background. 
Therefore, we exclude them from the potentially veto-triggering muons based on the beam time stamps.

In addition to the veto regions described in this section, HC-TFC also implements the Full Volume Vetoes described in \cref{sec:blind_cuts}.

	\subsection{Likelihood approach}\label{sec:mz_tfc}
		The implementation of the TFC described above applies hard cuts between muons, neutrons, and $\beta$-events to make a binary decision whether a candidate event should be regarded as a \Neu or \Cel event. 
In contrast, the approach detailed in this section tries a further optimization of tagging efficiency and dead-time reduction by assigning each event a non-binary likelihood to be cosmogenic. 
A likelihood ${\cal L} = \prod_i p_{i} $ is obtained from the product of several partial probabilities to be \Cel, $p_i$, that are based on a set of TFC discrimination parameters.

The discrimination parameters are largely the same as for the HC-TFC (\cref{sec:mi_tfc}), i.e.~the spatial distances and time elapsed with regards to the potential parent muons and neutrons. 
In addition, it permits to directly include the neutron multiplicity, the presence of neutrons within the MW, and the differential energy loss $dE/dx$ of the parent muon. 
The corresponding probability profiles $p_i$ can be obtained directly from data (see below) and are displayed in \cref{fig:probability_profiles}. 

The rationale behind this approach is that a \Cel candidate closely missing all hard-cut criteria to be cosmogenic would still score high in the likelihood approach, while an event fulfilling only narrowly one selection criterium will not be selected. 
Thus, the likelihood is expected to do particularly well in those cases. 
It also permits to balance between spatial and time distance between muons (or neutrons) and \Cel candidates, meaning effectively that wider spatial cuts are implemented in case of shorter time differences. 

A similar tagging scheme for the cosmogenic isotope \A{Li}{9} has been successfully applied in the Double-Chooz experiment \cite{theta13-doublechooz,Li9-doublechooz}.

\subsubsection{Data-driven probability profiles}\label{sec:paramters}

\begin{figure}
		\centering
		\includegraphics[width=1\linewidth]{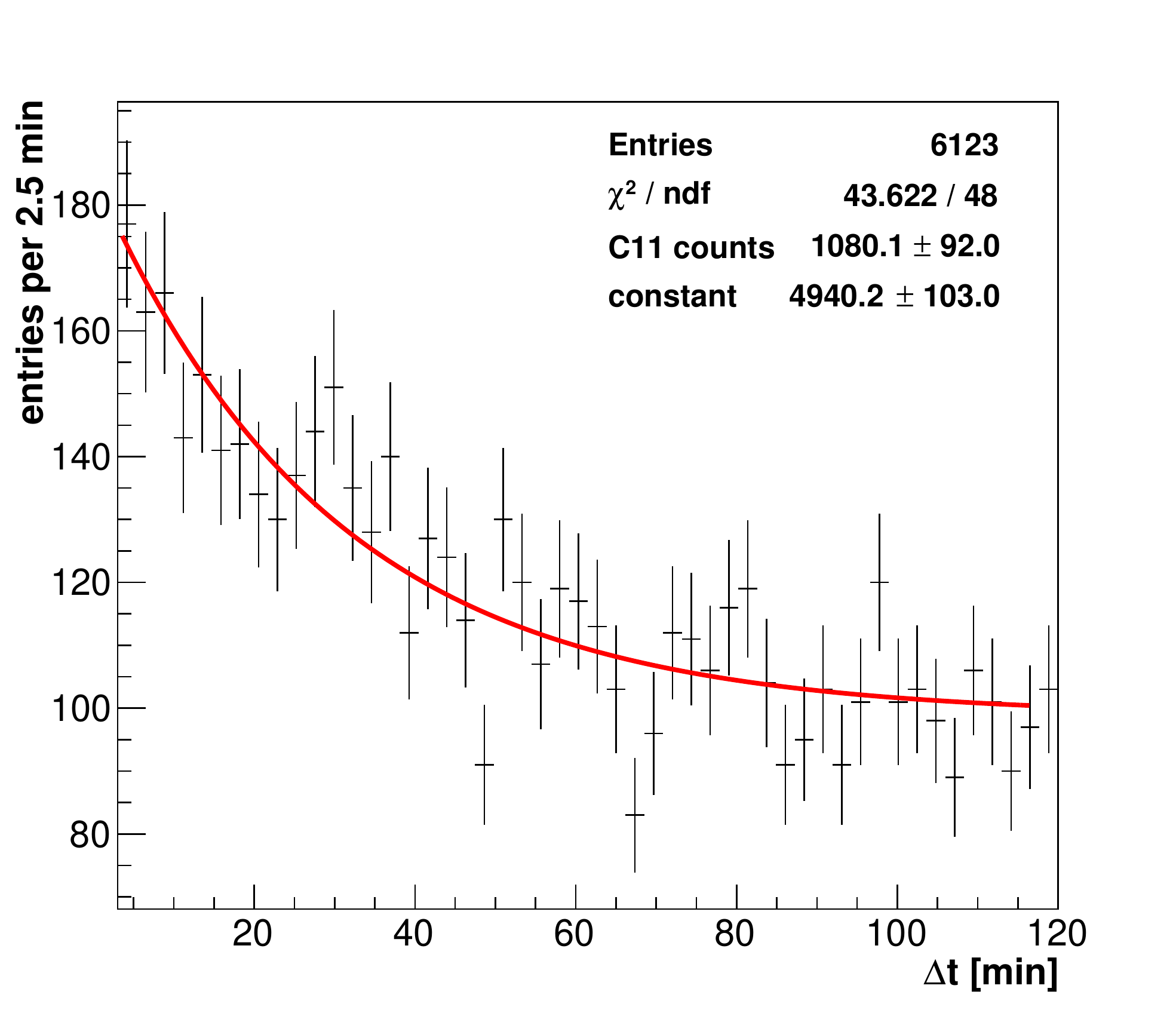}
		\caption{Example of a $\Delta t$ profile of 
		\Cel-candidates following muons. The distribution contains an exponentially decaying component from correlated muon-\Cel pairs as well as the flat component from accidental coincidences. In this special case, a lateral distance cut of $d_\mu = [1.0; 1.2)\text{ m}$ is applied. The ratio of correlated to uncorrelated pairs is used to produce the probability profiles for the LH-TFC tagging shown in Fig.~\ref{fig:probability_profiles}; in this case, the $6^{th}$ bin of histogram (b).}
		\label{fig:sample_dt_fit}
\end{figure}

The majority of probability profiles has been derived directly from the data, using the 2012 data set. 
For a given discrimination parameter $i$, the partial probability $p_i$ is determined by sorting the data set into several subsamples based on intervals of the parameter value. 
In each subsample, the relative contributions of  \Cel and uncorrelated events is evaluated based on the time difference ($\Delta t_{k}$) profile between a \Cel-candidate and all preceding muons $(k)$ in the last two hours. As an example, \cref{fig:sample_dt_fit} shows the $\Delta t_{k}$ profile of  \Cel-candidates selected by the distance from the parent muon track $d_\mu$ in the interval $[1.0\,{\rm m};1.2\,{\rm m})$. The $\Delta t_{k}$ distribution contains a flat component associated to accidental coincidences of unrelated $\mu$-candidate pairs as well as an exponentially decaying component (following the \Cel lifetime) of true $\mu$-\Cel coincidences. 
As shown in \cref{fig:sample_dt_fit}, the relative fraction of correlated \Cel in the subsample can be extracted by a fit containing those two components. This fraction is used as probability value $p_i$ in the construction of the partial probability profile. For each bin in a given profile, the corresponding data sample is selected, the $\Delta t_{k}$ distribution is fitted and the exponential fraction is determined. 

\begin{figure*}
		\centering
		\includegraphics[width=\linewidth]{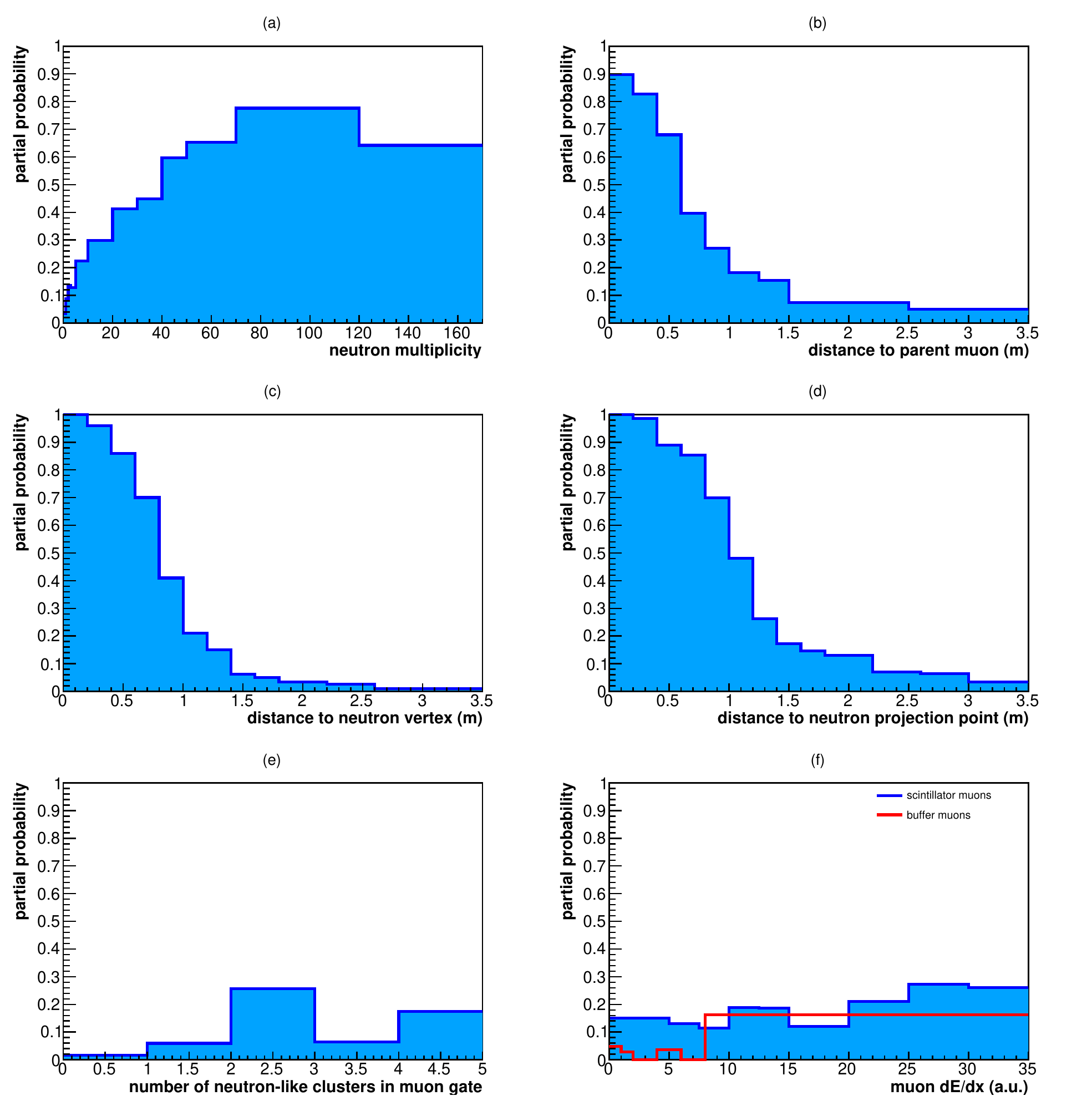}
		\caption{Overview of partial probability profiles of likelihood TFC discrimination parameters (description in the text).}
		\label{fig:probability_profiles}
\end{figure*}

We use a selection of seven discrimination parameters. The probability profiles obtained from the data for the first six are displayed in \cref{fig:probability_profiles}: 
\begin{itemize}
\item[\bf (a)] {\bf Neutron multiplicity, $p(n_n)$:} A higher number of neutrons, $n_n$, following a parent muon is a good indicator for hadronic shower formation. In some cases, a very luminous muon will flood the internal memories of the electronics, causing the NW to be entirely devoid of hits; based on a fit of the corresponding $\Delta t$-distribution (i.e.~selecting only muons followed by no hits in the subsequent NW), we determine for this case a value of $p(n_n)=0.885$, corresponding to a very high probability to find subsequent \Cel decays.  
\item[\bf (b)] {\bf Distance to the muon track, $p(d_\mu)$:} The probability to be \Cel is found to decline as function of $d_\mu$. 
In cases where muon track reconstruction fails, a probability $p(d_\mu)$ of 0.182 is assigned (based on the corresponding subsample).
\item[\bf (c)] {\bf Distance to the closest neutron vertex, $p(d_n)$:} Proximity to the neutron capture vertex was found to be a strong indicator for \Cel.
\item[\bf (d)] {\bf Distance to the closest neutron projection point, $p(d_{np})$:}  Proximity to the projection of the vertex on the muon track was identified as a similarly  efficient indicator.
\item[\bf (e)] {\bf Neutrons in the MW, $p(n_{\mu n})$:} While the entire MW in the wake of the initial muon signal is usually dominated by afterpulses, a special clustering algorithm has been devised to find superimposed peaks from cosmogenic isotopes and neutrons. 
While the vertex of these events cannot be reconstructed, the mere presence of these late clusters provides a (weak) indication of shower formation and hence \Cel production.
 \item[\bf (f)] {\bf $\mathbf{dE/dx}$ of the parent muon, $p(dE/dx)$:} High energy deposition per unit muon path length is a good indicator for a muon-induced hadronic shower. Here, we consider the visible energy of the muon per total path length as an observational estimator for the physical quantity. Two different distributions are used, one describing muons that have crossed the scintillator volume (scintillator muons), the other for those passing it by (buffer muons).
 \item[\bf (g)] {\bf Time elapsed since muon, $p(\Delta t)$:} The radioactive decay of \Cel means that a large time difference to the parent muon are less likely. The basic probability assigned is an exponential decay with the \Cel lifetime.
\end{itemize}
It is worthwhile to note that the partial profiles obtained do not constitute probability density functions in a mathematical sense. For instance, a good fraction of the accidental coincidences has been removed from the $\Delta t$ profiles by selecting only events inside the \Cel energy range and within a spherical volume of 3\,m radius to remove both low-energy radioactive decays and external backgrounds.
For some of the variables that only weakly select for \Cel (e.g.~the muon dE/dx), it was necessary to apply additional cuts to a second discrimination parameter to further enhance the signal-to-background ratio. Finally, the integral of the distributions shown in \cref{fig:probability_profiles} are not normalized. However, this is of no practical consequence for the construction of the likelihood since the contribution of the $p_i$ are {\it a posteriori} tuned by weight factors $w_i$ (see below).

\subsubsection{Construction of the likelihood}

\begin{table*}[t]
	\centering
	\begin{tabular}{llcccccccc}
		\hline
		$j$ 	&emphasis			& $q_j$	& $w(d_\mu)$	& $w(n_n)$ & $w(n_{\mu n})$ & $w(d_n)$ & $w(d_{np})$ & $w(\frac{dE}{dx})$ & $w(\Delta t$)\\
		\hline
		1	& closest $n$ vertex		& 0.025	&	1	&	1	&	0	&	2	&	0	&	0	&	1\\
		2 	& closest $n$ proj.		& 0.436	&	1	&	1	&	0	&	0	&	1	&	0	&	1\\
		3 	& neutrons in MW		& 0.740	&	1.5	&	0	&	1	&	0	&	0	&	1	&	1\\
		4 	& high $n$-mult.		& 0.370	&	1.5	&	1	&	0	&	0	&	0	&	1	&	1\\
		\hline
	\end{tabular}
	\caption{The case-dependent likelihood implementations $(j=1..4)$ for TFC (see text): Probability offsets $q_j$ and weight factors $w_{ij}$ are listed.}
	\label{tab:likelihoods}
\end{table*}

\noindent The construction of a single likelihood ${\cal L}$ for a given \Cel candidate necessitates the expansion of the simple likelihood definition to the expression
\begin{equation}\label{eq:likelihood2}
{\cal L}_{jk} = q_j \cdot \prod_i (p_{ik})^{w_{ij}},
\end{equation}
where $\prod_i p_{ik}$ is the product of the probability values derived from the profiles shown in Fig.~\ref{fig:probability_profiles}, with the index ($k$) denoting a specific parent muon. The meaning of the individual parameters $q_j$, $w_{ij}$ is explained below.

There are two considerations that drive us to building the likelihood in this way.
First, each candidate can be associated to a large number of preceding muons, i.e.~for each muon-candidate pair, an individual likelihood value ${\cal L}_k$ has to be computed. 
Second, the discrimination parameters feature different levels of indicative force for tagging \Cel. 
Moreover, some are statistically interdependent, some on the other hand contradictory: for instance, while the distances to the neutron vertex and projection, $p(d_n)$ and $p(d_{np})$, are good \Cel indicators on their own, their product is likely to be large only for candidates at similar distance from both and thus highly misleading for the purpose of \Cel tagging.

To address these issues, the partial probabilities $p_i$ are weighted by exponents $w_i$ that re-scale their relative contribution to the calculation of $\cal L$. Note that the introduction of the weights $w_i$ does preserve the shapes of the individual probability profiles ($p_i$) shown in Fig.\ \ref{fig:probability_profiles}.
In addition, we consider different implementations $(j)$ of the final likelihood, each one putting emphasis on a particular \Cel tagging strategy by re-weighing/excluding different sets of discrimination parameters ($i$):
 \begin{itemize}
\item[\bf (1)] {\bf Proximity to a neutron vertex} is given a special emphasis while the distance to the next projection point is not included.
\item[\bf (2)] {\bf Proximity to a neutron projection point} puts the tagging priority opposite of (1).
\item[\bf (3)] {\bf Presence of neutrons in the MW} bears the complication that no vertex reconstruction is available for neutrons in the MW. Thus, only the presence of neutrons and muon-related information is regarded.
\item[\bf (4)] {\bf Muons featuring high neutron-multiplicities} are usually very luminous, again impacting spatial reconstruction for the subsequent neutron vertices. Therefore, vertex information is not regarded for $n_n>20$, given instead larger emphasis to muon-related parameters.
\end{itemize}

The corresponding weights $w_{ij}$ are listed in \cref{tab:likelihoods} and are set to 0 if a specific $p_i$ is not to contribute to the likelihood.

Subsequently, the different implementations $(j)$ are evaluated and the variant providing the highest likelihood value ${\cal L}_{jk}$ is selected. To permit the direct comparison of the ${\cal L}_{jk}$, we have defined for each variant $(j)$ an optimum working point for what regards \Cel tagging efficiency and \Cel-depleted exposure. These figures of merit are introduced in \cref{sec:comparisons}. In order to compensate the differences in absolute scale, we apply {\it a posteriori} a scaling factor $q_j$ to each likelihood ${\cal L}_{jk}$, permitting a direct comparison between the four implementations $(j)$.

This procedure provides a single likelihood value ${\cal L}_k$ for a specific $\mu$-candidate pair. Also the likelihoods obtained for all possible parent muons $(k)$ are compared and again the maximum value is chosen. Thus, the final likelihood ${\cal L}$ for a specific \Cel candidate is
\begin{equation}\label{eq:likelihood3}
{\cal L} = {\rm max}_k \left[ {\rm max}_j {\cal L}_{jk} \right].
\end{equation}
All numerical values of the parameters listed in \cref{tab:likelihoods} have been optimized using the 2012 data set regarding tagging efficiency and preserved exposure (see \cref{sec:comparisons}).
Like the partial probabilities $p_i$, the resulting $\cal L$ thus cannot be considered a likelihood in a strict mathematical sense. However, a high value of $\cal L$ is indeed a reliable indicator for the cosmogenic origin of an event. 

\subsubsection{Application as \Cel event classifier}

\begin{figure}
		\centering
		\includegraphics[width=1\linewidth]{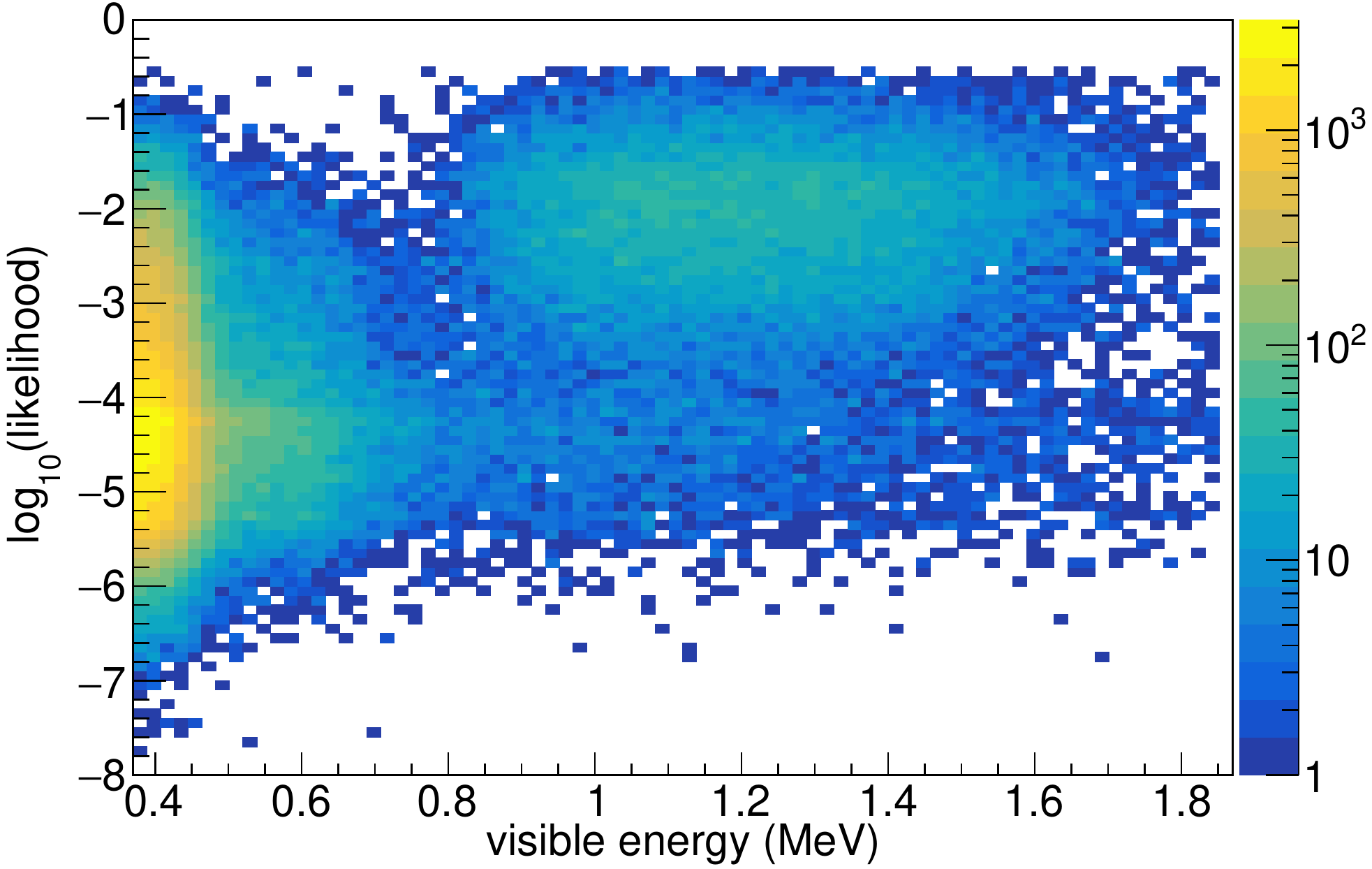}
		\caption{Distribution of TFC likelihood ${\cal L}$ as a function of visible energy. Two event populations can be identified. The upper band centered around ${\cal L}\sim10^{-2}$ are \Cel decays concentrated in the \Cel energy range. Contrariwise, the lower band centered on ${\cal L}\sim10^{-4}$ consists of \Neu events. It is worth to note that the large spectral excess of \Neu events at low energies only seems to suggest a vertical structure while it is in fact still centered around the same average value of ${\cal L}\sim10^{-4}$. Based on the 2012 data set.}
		\label{fig:likelihood_distribution}
\end{figure}

Differently from the HC-TFC, the LH-TFC provides the likelihood as a continuous variable for classification as a \Cel or \Neu event. The distribution of $\cal L$ as a function of the visible energy is shown in  \cref{fig:likelihood_distribution}. Indeed, two event categories are discernible: a \Neu band centered slightly below ${\cal L}\sim10^{-4}$ and stretching over all visible energies; and a smaller cloud of events centered on ${\cal L}\sim10^{-2}$ and concentrated in the energy range 0.75--1.87~MeV that can be associated with \Cel events.
This is a demonstration that the likelihood ${\cal L}$ is indeed a potent classifier for \Cel events. Depending on the application, regions in this plot can be selected to either obtain a very pure \Cel or \Neu sample. The standard application for neutrino analyses is the introduction of an energy-independent cut at ${\cal L}_0=10^{-4}$ that removes the majority of the \Cel events, while keeping about 70\% of the \Neu events. Analogously to the HC-TFC, this fraction is somewhat reduced when the Full Volume Vetoes discussed in the next section are applied in parallel. 

	\subsection{Full Volume Regions}\label{sec:blind_cuts}
		There are two situations in which the information on the parent muon and neutron events may be lost: during run breaks and during saturating muon events. 
These lead both algorithms to apply Full Volume Vetoes (FVV) to effectively reject potential \Cel events.
We briefly describe them here. 
\vspace{-3mm}
\paragraph{Run Breaks.} The Borexino data taking is organized in runs with a programmed duration of 6\,h although they can be shorter if automatically or manually restarted upon problems. 
The gap between runs can vary from a few seconds (ordinary restart) to one hour (weekly calibration) or exceptionally to a few days in case of major problems (blackouts, hardware failures).
The gap between runs is a time where unrecorded muons could potentially generate \Cel in the scintillator which later decays after the run is restarted. 
TFC algorithms apply a FVV called After-Run-Break (ARB) veto: all events at the beginning of each run are tagged as potential \Cel without looking for the parenthood of muon-neutron pairs.
The duration of the ARB veto is generally taken to be $\Delta t_{\textup{ARB}}=10 + 60 \cdot ( 1-e^{-3\Delta t_{\textup{gap}}})$\,min, where $\Delta t_{\textup{gap}}$ is the duration of the gap to the end of the previous run in minutes. 
\vspace{-3mm}
\paragraph{Showering muons.}
Muons crossing several meters of scintillator and/or inducing hadronic showers are potentially generating multiple \Cel along the track. 
At the same time the high amount of light can provoke the saturation effects described in \cref{sec:borexino}. 
This affects the position reconstruction of subsequent neutrons and can reduce the neutron detection efficiency in very extreme cases. Therefore, both algorithms apply a FVV after events with number of empty boards $N_{\rm EB}>N^{\rm thr}_{\rm EB}$ and $n_{n}>0$ for 4--5 $\tau_{^{11}\rm C}$.
We observe that the average value of $N_{\rm EB}$ for non saturating events is slowly changing through the data set, mostly increasing due to the diminishing number of live channels. 
In addition a few rearrangement of the electronics and cabling also impacted on this parameter. 
In order to take this into account we adjust $N^{\rm thr}_{\rm EB}$ to a value above the bulk of the distribution in each subset of data defined by the same electronics configuration and let the outliers trigger the FVV.

\vspace{3mm}
It should be noted that the data within FVV are considered when TFC algorithms search for muon-neutron pairs.

FVV are very important to keep the tag efficiency high but have a high price in terms of exposure.
This consideration led to the development of the alternative BI algorithm described in \cref{sec:bursts}.

	\subsection{TFC Comparison}\label{sec:comparisons}
		We present the performance of the two TFC algorithms described in terms of two desiderata. 
The TFC technqiue aims to maximize the \Cel tagging efficiency in the \Cel-enriched data set, while retaining the maximum exposure fraction in the \Cel-depleted set. 
The trade-off between those two competing requirements is ultimately decided by several parameters for the HC-TFC and by a single threshold ${\cal L}_0$ in case of the LH-TFC.
While the working point we present here is the result of a careful optimization, the best parameter combinations for each neutrino analysis can be determined independently and may be to some extent different than what is presented here.

\paragraph{Exposure fraction.} The first figure of merit is the fraction of the exposure $\varepsilon_\text{depleted}$ that remains in the \Cel-depleted data set after TFC application. As laid out in \cref{sec:tfc}, $\varepsilon_\text{depleted}$ is determined statistically by inserting random vertices (both position and time) into the experimental data stream and counting the fraction that is selected by the TFC criteria, i.e.~would be included in the \Cel-depleted data set.

\paragraph{Tagging efficiency.} The second figure of merit is the tagging efficiency $e_\text{tagged}$. Its value is estimated based on a relatively narrow energy range from 1.2 to 1.4\,MeV that is selected to minimize the presence of all spectral components but \Cel. 
In order to correct for the loss of live exposure in the depleted data set, $e_\text{tagged}$ is defined as:
\begin{equation}
	e_\text{tagged} = 1 - \cfrac{\text{\Cel}_\text{untagged}}{\text{\Cel}_\text{total}}\,\cdot\cfrac{1}{\varepsilon_\text{depleted}}
	\label{eq:taggingpower}
\end{equation}
where $\text{\Cel}_\text{total}$ and $\text{\Cel}_\text{untagged}$ are the total number of events  and that of untagged events, respectively. It is worth noting that $e_\text{tagged}$ is actually a lower limit, as spectral analysis suggest a presence of $\sim$5\% of non-\Cel events in the evaluated energy range.

\begin{figure}[ht!]
	\centering
		\includegraphics[width=1\linewidth]{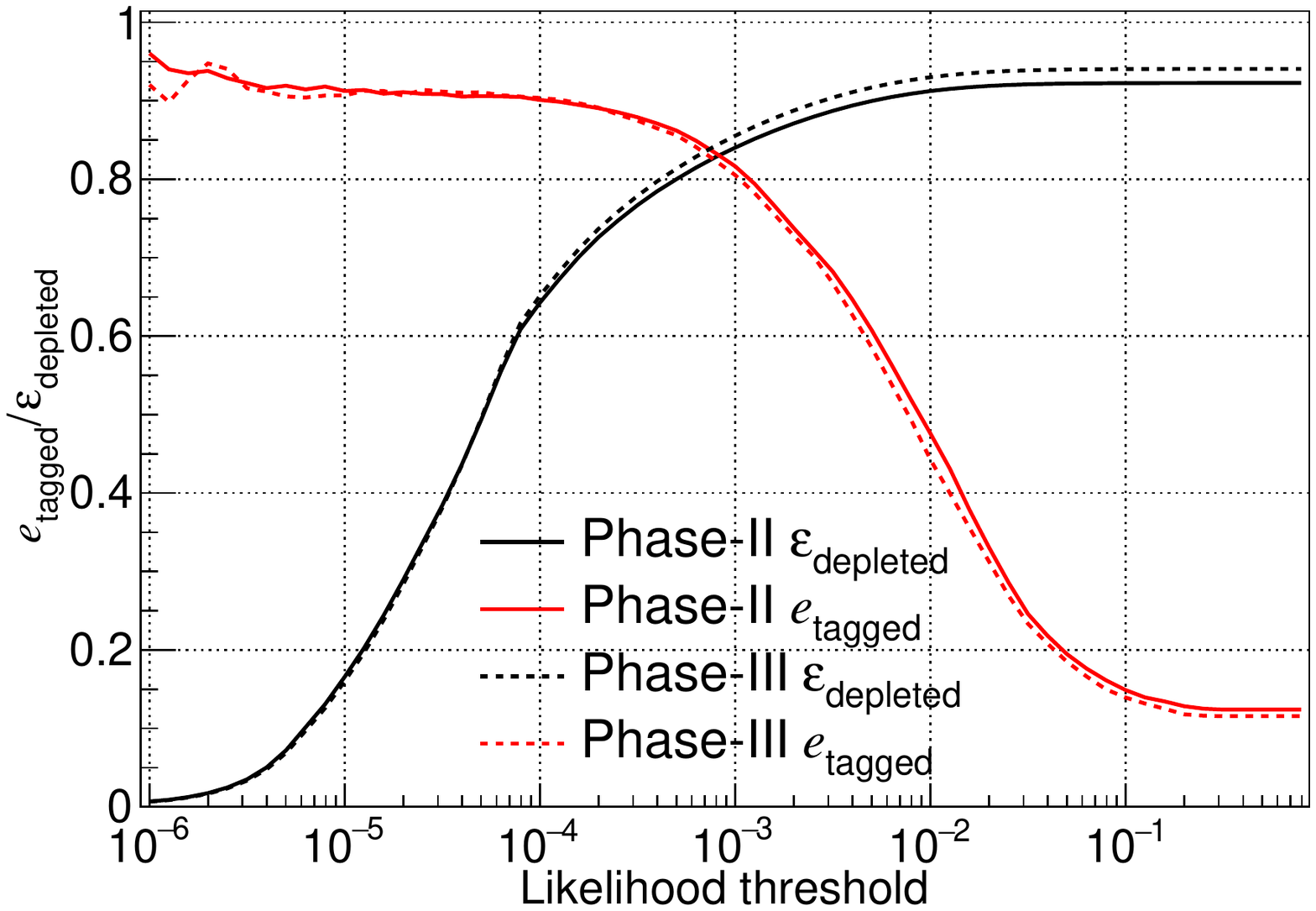}
		\caption{Performance of the LH-TFC as a function of its threshold value ${\cal L}_0$. \Cel tagging efficiency ($e_\text{tagged}$) and the exposure of the \Cel-depleted data set ($\varepsilon_\text{depleted}$) can be adjusted to the need of a particular analysis. For solar analysis, the default choice is ${\cal L}_0=10^{-4}$.}
		\label{fig:likelihood_efficiency_exposure}
\end{figure}

\paragraph{TFC tuning.} For both TFC algorithms a large flexibility of performance can be obtained tuning the selection parameters. In the case of the LH-TFC this is particularly easy as tagging efficiency and exposure can be varied as a function of the likelihood threshold ${\cal L}_0$ as shown in \cref{fig:likelihood_efficiency_exposure}. 
While a value of ${\cal L}_0=10^{-4}$ provides high tagging power for a moderate exposure loss that is ideal for solar neutrino analyses, a choice of ${\cal L}_0=10^{-2}$ would mean that a much larger fraction of the exposure would be preserved while only half of the \Cel candidates are tagged. In this case, the corresponding \Cel-enriched sample would almost exclusively consist of real \Cel decay events.

\paragraph{TFC performance.} The performance results are presented in \cref{tab:TFCcomparison} and are based on the data of Borexino Phase-II and Phase-III introduced in \cref{sec:borexino}. 
Although the performance is essentially independent from the choice of the fiducial volume, for consistency we use the same fiducial volume of the $pep$ and CNO neutrino analyses (\cref{sec:borexino}).

\begin{table*}[t]
	\centering
	\begin{tabular}{ll|cc}
							&						& Phase-II	& Phase-III\\
		\hline
		\multirow{2}{*}{Hard Cut}	& Tagging efficiency $(e_\text{tagged})$		& 90.2\%	& 90.7\%	\\
							&Exposure fraction $(\varepsilon_\text{depleted})$	& 63.3\%	& 63.6\%	\\
		\hline
		\multirow{2}{*}{Likelihood}	& Tagging efficiency $(e_\text{tagged})$		& 89.8\%	& 90.1\%	\\
							&Exposure fraction $(\varepsilon_\text{depleted})$	& 64.7\%	& 65.6\%	\\
		\hline
	\end{tabular}

	\caption{Tagging efficiency (calculated with \cref{eq:taggingpower}) and exposure fraction for the two TFC approaches. For the LH-TFC, ${\cal L}_0$ is set at $10^{-4}$.
	The statistical uncertainty on all values of $e_\text{tagged}$ is 0.5\%, while it is negligible on $\varepsilon_\text{depleted}$ values.}
	\label{tab:TFCcomparison}
\end{table*}

Both algorithms return a tagging efficiency of about 90\% while providing a \Cel-depleted exposure fraction of $\sim$64\%. While HC-TFC provides slightly better tagging, LH-TFC saves somewhat more exposure. 
In recent solar neutrino analyses\cite{bx_nusol}, both algorithms have been applied separately to investigate potential differences in the corresponding spectral fit results and to assess systematic uncertainties. 

\begin{figure*}[t]
	\begin{minipage}[t]{0.49\linewidth}
		\centering
		\includegraphics[width=\linewidth]{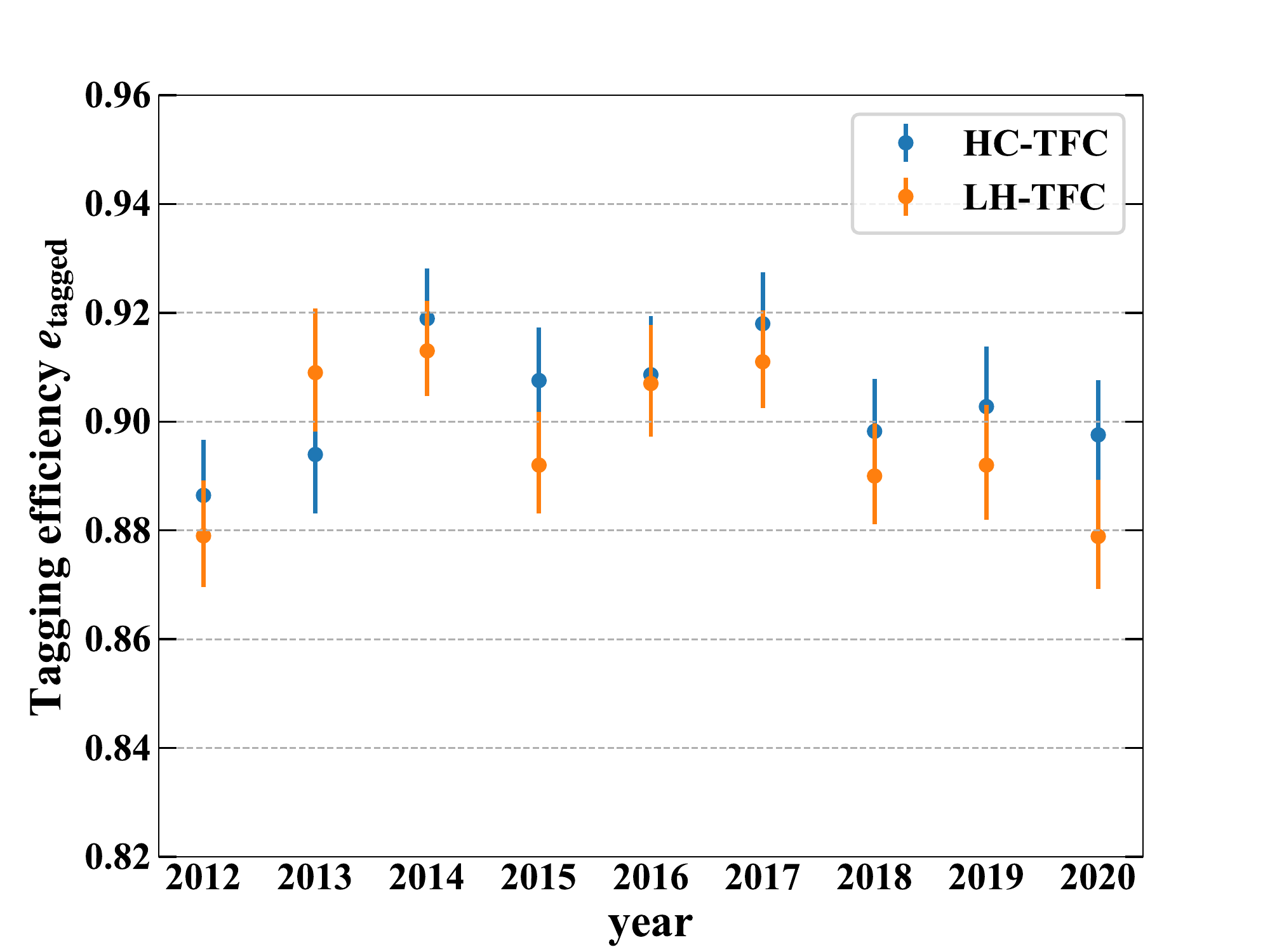}
		\caption{Tagging efficiency for HC-TFC (blue dots) and LH-TFC (orange dots) year by year.}
		\label{fig:stability_rejection}
	\end{minipage}
	\hfill
	\begin{minipage}[t]{0.49\linewidth}
		\centering
		\includegraphics[width=\linewidth]{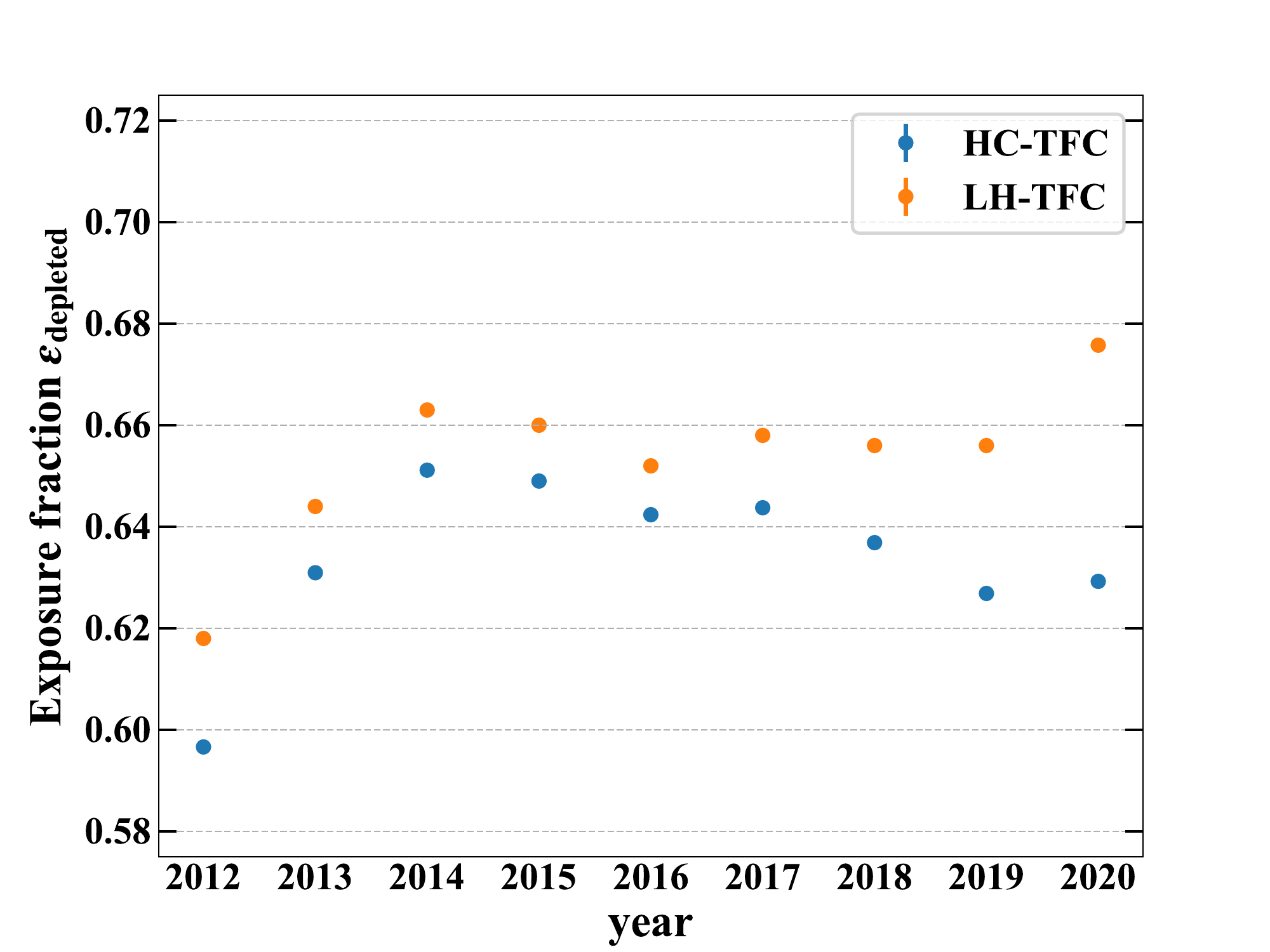}
		\caption{Exposure fraction of the depleted sample for HC-TFC (blue dots) and LH-TFC (orange dots) year by year.}
		\label{fig:stability_exposure}
	\end{minipage}
\end{figure*}

\paragraph{Time stability.} We also investigated the stability of the TFC performance on a yearly bases throughout Borexino phase-II and III and extending till the end of 2020.
In \cref{fig:stability_exposure,fig:stability_rejection} the \Cel tagging efficiency and the exposure fraction are shown for both TFC implementations. 
We attribute the slightly better performance of both TFC methods during Phase-III to a more continuous data taking which implies a lower impact of ARB cuts.
Moreover, we observe a drift in LH-TFC performance during Phase-III, trading slightly lower tagging efficiencies for higher exposure fractions compared to HC-TFC. In principle, this trend could be compensated by a year-by-year adjustment of the likelihood threshold ${\cal L}_0$. 

\section{\bfCel burst Identification}\label{sec:bursts}
	The implementations of the TFC discussed so far have been extensively tested and validated and were part of several of the Borexino solar-$\nu$ analyses. 

Here, we describe a novel method that further improves the identification of \Cel background events in Borexino and is potentially applicable as well to other large volume liquid scintillator detectors.

Most of \Cel is produced not by primary muon spallation but by secondary processes like $(p,n)$ reactions occurring in muon-induced hadronic showers. 
As a consequence, multiple \Cel nuclei are produced simultaneously in the detector and subsequently decay over a couple of hours. 
Consequently, identification of a \burst of possible \Cel candidates correlated in space (i.e.~along a hypothetical muon track) and time (several \Cel lifetimes) can provide a tag independent of any information on parent muon and neutron events.
This is of particular interest since this Burst Identification (BI) tag will be available even if the initiating muon and/or subsequent neutron captures have been missed or misidentified by the data acquisition.
Therefore, when applied in combination with either of the two TFC algorithms, the BI tag offers a more convenient alternative to the Full Volume Veto (FVV) introduced in \cref{sec:blind_cuts}. 

The BI algorithm presented here proceeds first to identify groups of \Cel candidates, i.e.~events in the correct energy window, with sufficient space and time correlation with each other.
Once a \burst has been identified, it is used to tag all events that show a correlation with the events that defined it.

\begin{figure*}[t]
	\centering
	\subfigure[Probability distribution from \Cel decay time in a $4\tau$ time window.]{\includegraphics[width=0.49\textwidth,trim= 0mm 0mm 10mm 12mm,clip]{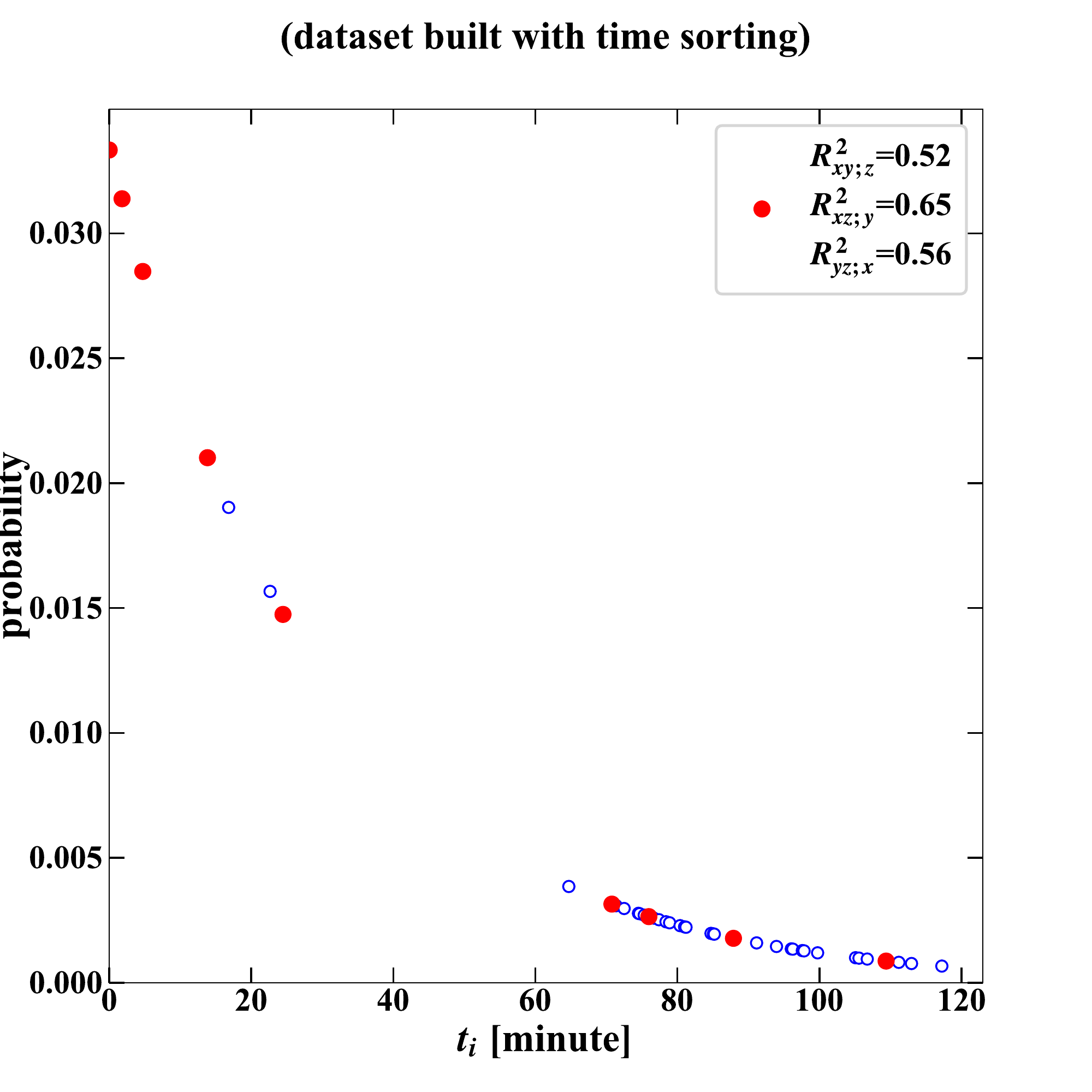}}
	\label{fig:burst_timepdf}
	\hfill
	\subfigure[Spatial distribution in the $xy$-plane.]{\includegraphics[width=0.49\textwidth,trim= 0mm 0mm 10mm 12mm,clip]{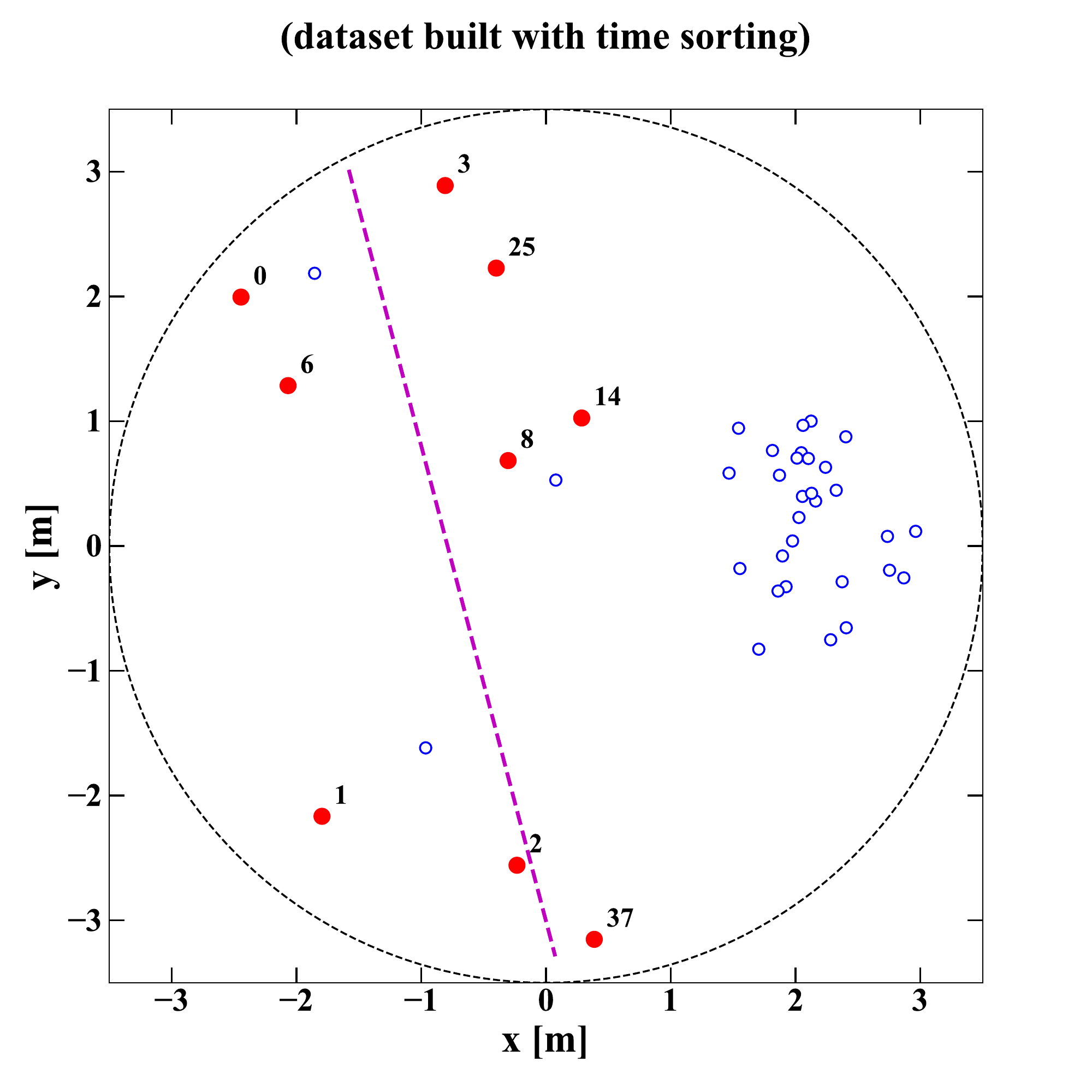}}
	\label{fig:burst_spacecorr}
	\caption{Sample group of events considered for building a \burst.
	Red full dots highlight the events that are finally selected as spatially correlated, with the multi-correlation factors $R^2$ reported in the legend in figure (a). 
	The numbers in figure (b) reflect the time ordering (for clarity only some are shown).
	The dashed circle shows the radius of 3.5\,m used to pre-select \burst candidates.
 	The dashed magenta line is the projection of the regression line. 
 	Open blue dots represent events not correlated with the first \burst. Note that both panels show a late and spatially dense population belonging most likely to a second coinciding \Cel{} \burst that is distinguished as separate by the algorithm.}
	\label{fig:burst}
\end{figure*}

\subsection{Step 1: finding the \burst}\label{sec:burst_step1}
\label{sec:burst_finding}

\paragraph{Time correlation.} The algorithm starts considering successively all \Cel candidates searching for time correlations. 
Candidate events are defined by a visible energy within 0.75--1.87\,MeV corresponding to the \Cel spectrum.
At this stage a very loose fiducial volume cut is applied: all events within a 3.5\,m radius are considered. 
This has been optimized for excluding most of the external background events, without loosing many \Cel events that could be helpful in finding a \burst. 
For each candidate, a subsequent time window corresponding to $4\tau_{^{11}\rm C}$ ($\sim$\,2 hours) is opened. 
In order to form a potential \burst, at least four \Cel candidates have to be present inside the window (including the initial one).
\cref{fig:burst} (a) shows a sample of events considered in the definition of a \burst to whom we have associated a probability based on the exponential decay time of \Cel.
We build a cumulative probability summing these values for all candidates, $\sum_i 1/\tau_{^{11}\rm C} \cdot \exp(-(t_i-t_0)/\tau_{^{11}\rm C})$, that is required to exceed a given threshold $p_\text{th}$. 
In the case of a run interruption within the observation window, this threshold is adjusted to take into account the dead time. 

In some rare cases, it may happen that a \burst falls within the selection window of a preceding unrelated \Cel candidate.
The latter can then be mistaken as the first event and seriously distort the construction of the \burst. 
In order to prevent this, we check whether the time distribution of events inside the selection window follows an exponential decay law. Specifically, if the time gap between first and second candidate is disproportionately large, the first event is considered unrelated to the \burst and discarded.

\begin{figure*}
	\centering
	\includegraphics[width=\linewidth]{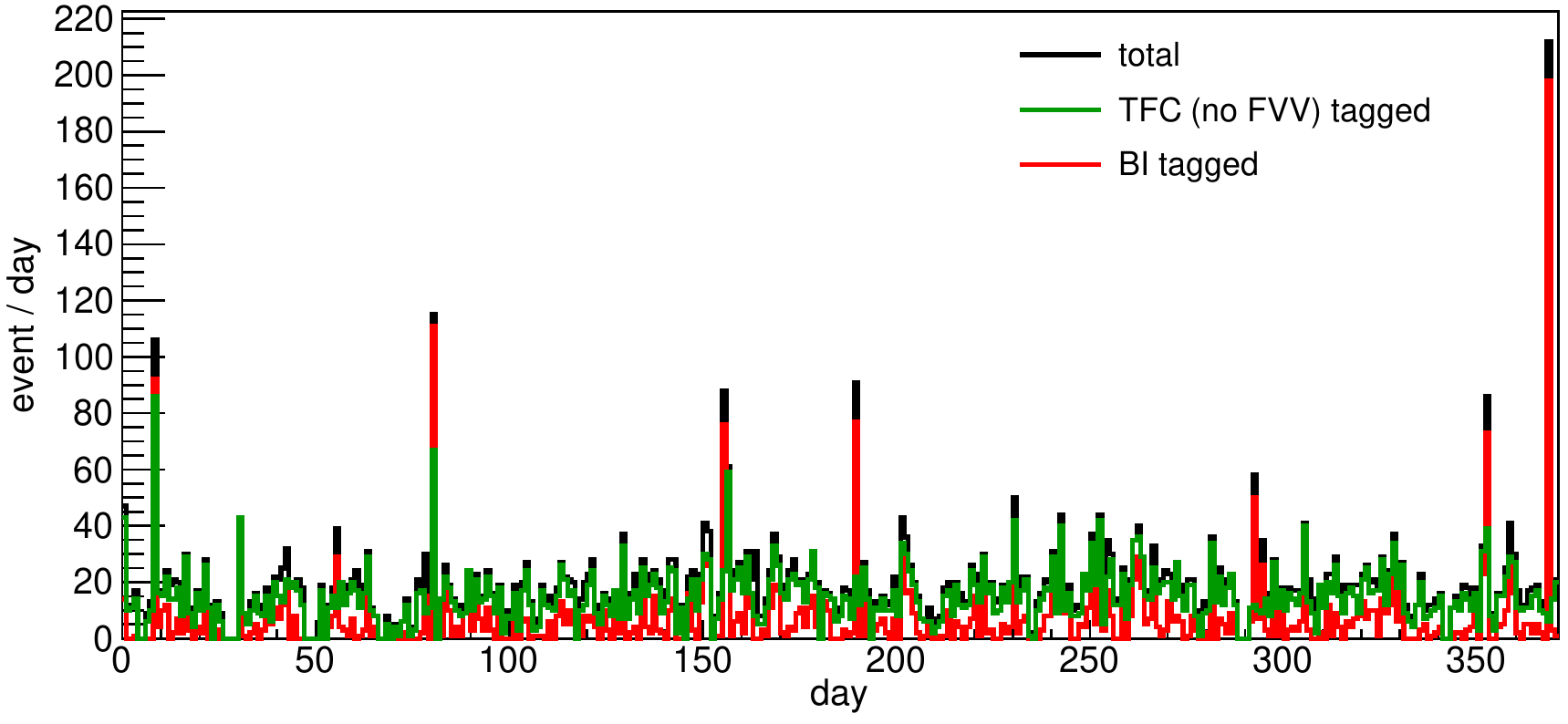}
	\caption{Events per day in the \Cel region inside the fiducial volume, in a year-long window of Phase-III.
	Black: total daily rate. Green: rate of events tagged by the TFC with the Full Volume Veto (FVV) switched off (see text for details). Red: rate of events tagged by the Burst Identification (BI) algorithm. Spikes in the rate correspond to {\it bursts} of \Cel.}
	\label{fig:one_year_of_tag}
\end{figure*}

\paragraph{Spatial correlation.} 
After a time correlation is established, the algorithm proceeds to check if any spatial correlation exists among the time-selected \Cel candidates.
The basic assumption is that \Cel is produced and decays along a straight line (i.e.~a muon track). 
Events that are far from this line are likely to be accidental time coincidences and must not be included in the \burst. 
The hypothesis is tested via the correlation coefficients of the events coordinates upon projecting on the coordinate planes. 

\cref{fig:burst} (b) shows the projection on the $(xy)$-plane of the sample events.
Since the line is a freely oriented three-dimensional object, the use of classic Pearson correlation coefficients $\rho_{ij} = \text{cov}(i,j)/\sigma_i\sigma_j$ is not sufficient. 
Instead, the correlation is checked based on the three multi-correlation factors

\begin{equation}
	R^2_{ij;k} = (\rho_{ik}^2+\rho_{jk}^2 - 2\rho_{ij}\rho_{ik}\rho_{jk})/(1-\rho_{ij}^2)\text{,}
	\label{eq:multicorr}
\end{equation}
with $(i,j,k)$ denoting the different combinations of spatial coordinates $(x,y,z)$ that correspond to the coordinate planes.

Starting values for the $R^2_{ij;k}$ are computed based on the first four time-selected events (denoted as 0 to 3 in \cref{fig:burst}). 
For each following event, the $R^2_{ij;k}$ values are re-computed after adding the event to the \burst.
The event is considered aligned with the track if: 
(1) all new $R^2_{ij;k}$ values remain larger than a given lower limit $R^2_\text{lim}$; 
(2) the two largest new $R^2_{ij;k}$ values do not differ from the previous values by more than a tolerance margin $R^2_\text{tol}$.
Upon meeting both conditions the event is accepted to belong to the \burst and the new $R^2_{ij;k}$ values become the new reference. 
In \cref{fig:burst} the events finally included in the \burst after spatial selection are indicated by red filled circles. 
The final correlation factors $R^2_{{\rm burst};k}$ provide the thresholds for the application of the tagging (\cref{sec:burst_tagging}).
The procedure is repeated upon sorting the events after their distance to the initial candidate and the case showing a stronger correlation is preferred.

For the performance evaluation presented in \cref{sec:burst_performance}, the values of the parameters discussed above were chosen to be $p_\text{th} = 0.12$, $R^2_\text{lim} = 0.1$, and $R^2_\text{tol} = 40\%$ (20\%) for Phase-II (Phase-III). 
Although these parameter values are the result of a careful calibration of the algorithm, a different choice could be made in the context of a specific neutrino analysis in order to achieve a different trade-off between tagging efficiency and exposure fraction.

\begin{table*}[t]
	\centering
	\begin{tabular}{ll|cc}
													HC-TFC	&			& Phase-II	& Phase-III\\
		\hline
		\multirow{2}{*}{no FVV}				    & Tagging efficiency $(e_\text{tagged})$ 		& 79.5\%	& 77.6\%	\\
											    & Exposure fraction $(\varepsilon_\text{depleted})$	& 70.8\%	& 69.7\%	\\
		\hline
		\multirow{2}{*}{no FVV + BI}		    & Tagging efficiency $(e_\text{tagged})$		& 89.3\%	& 90.4\%	\\
											    & Exposure fraction $(\varepsilon_\text{depleted})$	& 68.3\%	& 66.7\%	\\
		\hline
		\multirow{2}{*}{with FVV (standard)}    & Tagging efficiency $(e_\text{tagged})$		& 90.2\%	& 90.7\%	\\
											    & Exposure fraction $(\varepsilon_\text{depleted})$	& 63.3\%	& 63.6\%	\\
		
	\end{tabular}
	\caption{Performance of the HC-TFC without Full Volume Veto (FVV) before and after the combination with the Burst Identification (BI) tag, compared with the standard HC-TFC already shown in \cref{tab:TFCcomparison}. The statistical uncertainty on all values of $e_\text{tagged}$ is 0.5\%, while it is negligible on $\varepsilon_\text{depleted}$ values.}
	\label{tab:Bursts_comparison}
\end{table*}

\begin{figure*}[t]
	\begin{minipage}[t]{0.49\linewidth}
		\centering
		\includegraphics[width=\linewidth]{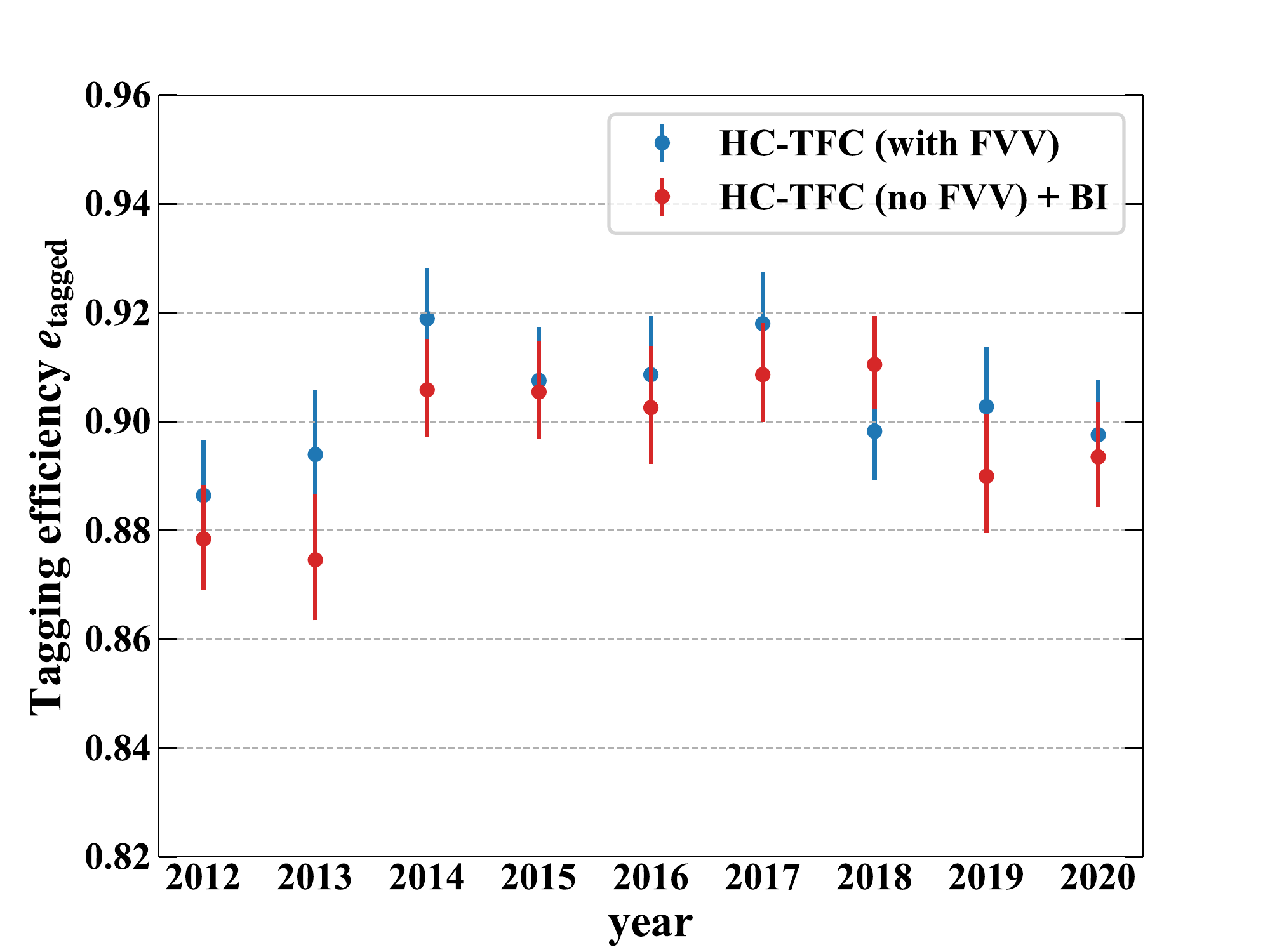}
		\caption{Tag efficiency year by year for TFC (with Full Volume Veto, FVV) and TFC (no FVV) + Burst Identification (BI), respectively in blue and red.}
		\label{fig:burst_stability_rejection}
	\end{minipage}
	\hfill
	\begin{minipage}[t]{0.49\linewidth}
		\centering
		\includegraphics[width=\linewidth]{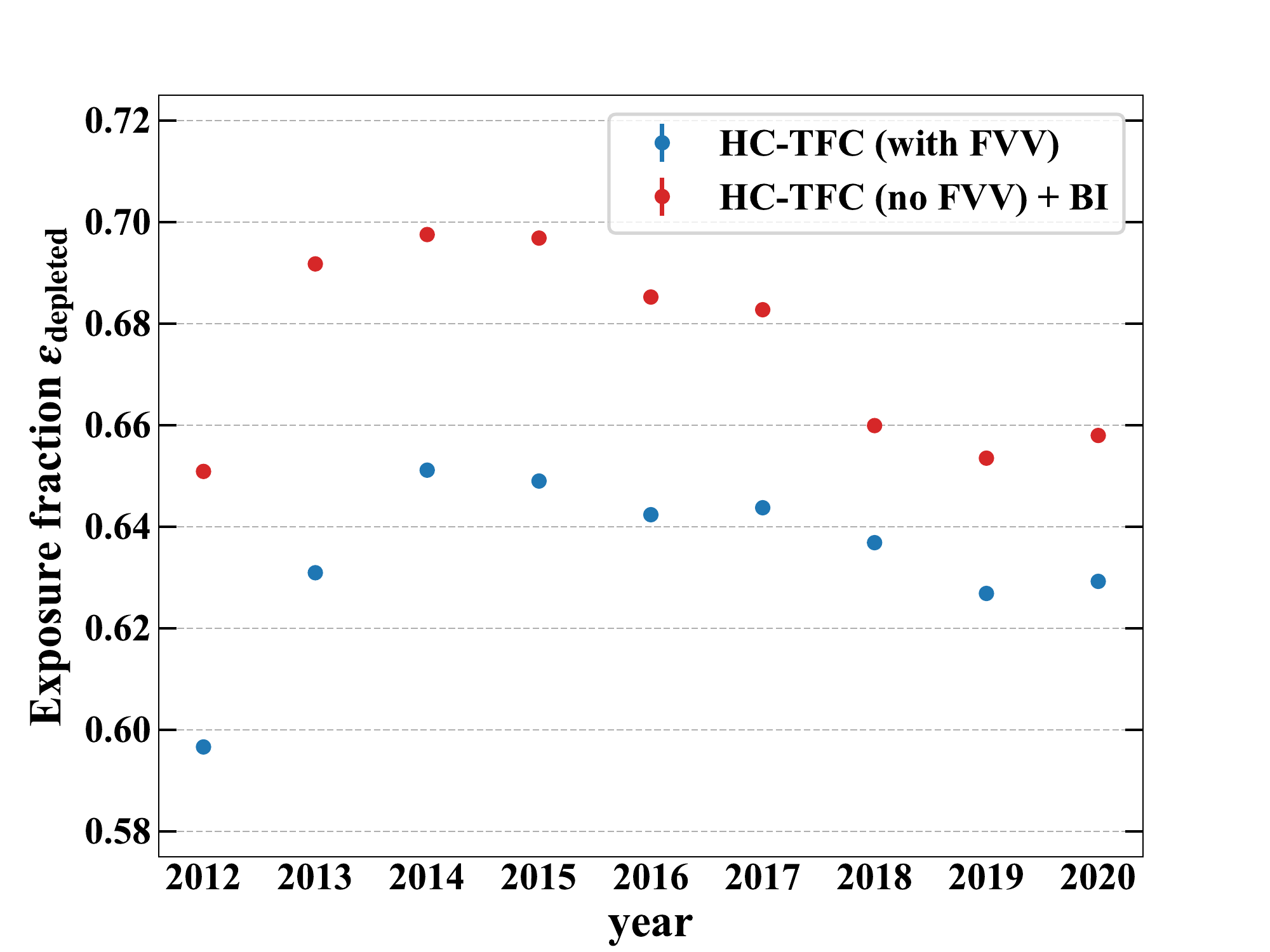}
		\caption{Exposure fraction of the depleted sample year by year for TFC (with FVV) and TFC (no FVV) + BI, respectively in blue and red.}
		\label{fig:burst_stability_exposure}
	\end{minipage}
\end{figure*}

\subsection{Step 2: tagging the \Cel}
\label{sec:burst_tagging}

Once a \burst is identified by the events selected as in \cref{sec:burst_finding}, all events within the \burst time window, regardless of their energy, are checked for their affiliation with the \burst and tagged as \Cel.  

The track underlying the \burst is found by a Theil-Sen regression\cite{theil,sen} returning the direction and the impact parameter based on the median of the tracks passing through all possible pairs of events of the \burst (this method is less influenced by outliers compared to those based on the mean). 
In \cref{fig:burst} (b), the projection of the track on the $(xy)$-plane is shown as a magenta dashed line.

Thereafter, events within the \burst time window are ordered by their distance to the track. 
Starting with the four closest events and then subsequently adding further events one by one, the algorithm proceeds to calculate the multi-correlation factors as in \cref{eq:multicorr}.
Newly added events are considered correlated with the track as long as all three $R^2_{ij;k}$ remain above the corresponding $R^2_{{\rm burst};k}$. 
Events may be tested for correlation multiple times if they fall within the time window of more than one track.
An event is tagged as \Cel by the BI algorithm if it is correlated within at least one track.

Similarly to the TFC, the BI algorithm calculates the exposure fractions via toy Monte Carlo. 
Simulated events are uniformly distributed inside the fiducial volume and their correlation with all \burst tracks is computed. 
The fraction of events for which $R^2_{ij;k}<R^2_{{\rm burst};k}$ for all tracks corresponds to the live exposure fraction.

\subsection{BI Performance}
\label{sec:burst_performance}

As discussed in \cref{sec:blind_cuts}, the TFC algorithms can only identify \Cel decays for which the parent muon and subsequent neutrons have been detected. 
There are situations following breaks in data acquisition and especially very bright muon events saturating the electronics where this requirement is not satisfied. 
The Full Volume Vetoes (FVV) that have to be introduced to compensate for this deficiency are very costly in exposure for the \Cel-depleted sample.
Here we demonstrate that the same task is more efficiently fulfilled by the BI algorithm.

\cref{fig:one_year_of_tag} shows the daily rate of events in the \Cel energy range and inside the fiducial volume (\cref{sec:borexino}) over the course of a sample period of one year.
It also shows the rates of those events identified as \Cel by the HC-TFC without FVV and of those identified by the BI algorithm.
While this basic TFC identifies most of the \Cel events during regular data taking, it misses a larger fraction of events during the bursts due to exceptional high-multiplicity \Cel production. 
This behaviour is expected since \Cel-{\it bursts} are accompanied by severe saturation of the electronics.
On the other hand, the BI algorithm shows heightened efficiency for these peak regions. 

Hence, an event is assorted to the \Cel-enriched sample in case it is tagged either by the TFC (without FVV)\footnote{We chose HC-TFC as reference in this context, but the combination of LH-TFC and BI performs similarly.} or by the BI algorithm. 
Similarly, we evaluate the exposure based on a shared toy Monte Carlo generation. 
Results for the combined tagging efficiency and the \Cel-depleted exposure fraction are shown in \cref{tab:Bursts_comparison}. 
The combination shows a tagging efficiency similar to the full TFC (\ie including FVV), but as expected it preserves a $\sim$3-5\% larger exposure for the \Cel-depleted data set. 

Similarly to \cref{sec:comparisons}, we have studied the time stability of the combined tags. 
The results  are shown in \cref{fig:burst_stability_rejection,fig:burst_stability_exposure} for Phase-II, Phase-III, and extending until the end of 2020. 

\section{Conclusions}\label{sec:conclusions}
	Low-energy neutrino detection in organic liquid scintillators suffers from spallation of carbon nuclei by cosmic muons. 
We have presented the state-of-the-art tagging and veto techniques for the most frequent spallation product \Cel that have been developed in Borexino in the context of solar $pep$ and CNO neutrino analyses. 

We have described the Three-Fold Coincidence (TFC) technique that makes use of time and spatial correlation among parent muons, neutrons, and \Cel decay events. 
We have given details of its two current implementations, a Hard-Cut and a Likelihood-based approach.
Scanning the data of Borexino Phases II (2012-2016)  and III (2016-2020), we showed that both methods return stable results over a long time period. Typical tagging efficiencies exceed 90\%, while the \Cel-depleted data set retains about 65\% of the exposure.

Finally, we discussed the \Cel-{\it burst} tagging technique that exploits the fact that \Cel is often created in bursts by major spallation events. 
This novel approach searches for clusters of \Cel candidates that feature an exponentially decaying time behavior and are closely aligned along a track. 
We showed how this technique can be combined with the TFC and determined the corresponding improvement in surviving exposure fraction (66.7\,\% vs. 63.6\%) for essentially the same tagging power on Phase-III data.
Given the net benefit, the method is intended for use in future solar neutrino analyses of Borexino.

While native to Borexino, these methods can be easily adjusted to the identification and veto of other isotopes or to alternative experimental setups, and thus may be of benefit for a wide range of present and future low-energy neutrino and other low-background experiments.

\begin{acknowledgement}
	The Borexino program is made possible by funding from Istituto Nazionale di Fisica Nucleare (INFN) (Italy), National Science Foundation (NSF) (USA), Deutsche Forschungsgemeinschaft (DFG) and Helmholtz-Gemeinschaft (HGF) (Germany), Russian Foundation for Basic Research RFBR (Grant \mbox{19-02-00097 A}), RSF (Grant \mbox{21-12-00063}) (Russia), and Narodowe Centrum Nauki (NCN) (Grant No.\ UMO \mbox{2017/26/M/ST2/00915}) (Poland). This research was supported in part by PLGrid Infrastructure.
	We acknowledge the generous hospitality and support of the Laboratory Nazionali del Gran Sasso (Italy). 
\end{acknowledgement}


\bibliographystyle{spphys}
\bibliography{bibliography}

\end{document}